# One-step synthesis of graphene containing topological defects

Benedikt P. Klein[1,2†], Matthew A. Stoodley[1,2], Joel Deyerling[3], Luke A. Rochford[1,4], Dylan B. Morgan[2], David Hopkinson[1], Sam Sullivan-Allsop[5], Fulden Eratam[1], Lars Sattler[6], Sebastian M. Weber[6], Gerhard Hilt[6], Alexander Generalov[7], Alexei Preobrajenski[7], Thomas Liddy,[1,8] Leon B. S. Williams[1,9,10], Tien-Lin Lee[1], Alex Saywell[11], Roman Gorbachev[5], Sarah J. Haigh[5], Christopher Allen[1,12], Willi Auwärter[3], Reinhard J. Maurer[2,13*] and David A. Duncan[1*‡]

[1]Diamond Light Source, Harwell Science and Innovation Campus, Didcot, OX11 0DE, United Kingdom.
[2]Department of Chemistry, University of Warwick, Gibbet Hill Road, Coventry, CV4 7AL, United Kingdom.
[3]Physics Department E20, TUM School of Natural Sciences, Technical University of Munich, James-Franck-Straße 1, 85748 Garching, Germany.
[4]Department of Earth Sciences, University of Cambridge, Downing Street, Cambridge, CB2 3EQ, United Kingdom.
[5]National Graphene Institute, University of Manchester, Oxford Road, Manchester, M13 9PL, United Kingdom.
[6]Institute of Chemistry, Carl von Ossietzky University Oldenburg, Carl-von-Ossietzky-Straße 9-11, 26111 Oldenburg, Germany.
[7]MAX IV Laboratory, University of Lund, Fotongatan 2, 224 84 Lund, Sweden
[8]School of Chemistry, University of Nottingham, University Park, Nottingham, NG7 2RD, United Kingdom
[9]School of Chemistry, University of Glasgow, University Avenue, Glasgow, G12 8QQ, United Kingdom.
[10]School of Physics & Astronomy, University of Glasgow, University Avenue, Glasgow, G12 8QQ, United Kingdom.
[11]School of Physics & Astronomy, University of Nottingham, University Park, Nottingham, NG7 2RD, United Kingdom
[12]Department of Materials, University of Oxford, Parks Road, Oxford, OX1 3PH, United Kingdom.
[13]Department of Physics, University of Warwick, Gibbet Hill Road, Coventry, CV4 7AL, United Kingdom.
*Corresponding authors. E-mails: r.maurer@warwick.ac.uk; david.duncan@nottingham.ac.uk;
†Current address: Research Center for Materials Analysis, Korea Basic Science Institute, 169-148 Gwahak-ro, Yuseong-gu, Daejeon 34133, Republic of Korea
‡ Current address: School of Chemistry, University of Nottingham, University Park, Nottingham NG7 2RD, United Kingdom

**Abstract:** Chemical vapour deposition enables large-domain growth of ideal graphene, yet many applications of graphene require the controlled inclusion of specific defects. We present a one-step chemical vapour deposition procedure aimed at retaining the precursor topology when incorporated into the grown carbonaceous film. When azupyrene, the molecular analogue of the Stone-Wales defect in graphene, is used as a precursor, carbonaceous monolayers with a range of morphologies are produced as a function of the copper substrate growth temperature. The higher the substrate temperature during deposition, the closer the resulting monolayer is to ideal graphene. Analysis, with a set of complementary materials characterisation techniques, reveals morphological changes closely correlated with changes in the atomic adsorption heights, network topology, and concentration of 5-/7-membered carbon rings. The engineered defective carbon monolayers can be transferred to different substrates, potentially enabling applications in nanoelectronics, sensorics, and catalysis.

**Main Text:**

Chemical vapour deposition (CVD) procedures have realised single-domain growth of ideal graphene on the meter scale.[1] However, for many applications, notably as a catalytic support material,[2] a battery electrode,[3] a gas sensor[4,5] or an electronic component,[6] perfectly crystalline graphene is not an ideal material. The inclusion of defects into graphene is often necessary to improve specificity of binding to graphene, for catalytic or gas sensor applications, or to modify graphene's electronic and magnetic properties for nanoelectronics or valleytronics applications.[2-6] For example, a Stone-Wales defect, which replaces four 6-membered rings with two 5-

and two 7-membered rings, increases the strength of interaction between graphene and its substrate;[7, 8] vacancy and heteroatom defects in graphene can introduce magnetic order at finite temperature.[9]

Current methods for including defects in graphene are, generally, post-processing modifications of ideal graphene or graphene oxide,[5, 10, 11] though these methods lack control and have poor defect homogeneity.[12] Ullman coupling methods[13] can yield highly crystalline films. However, Ullman coupling requires halide groups that typically contaminate the product[14] and can have a deleterious effect on the film quality at elevated temperatures.[15] Contamination and defect inhomogeneity will hinder studies into elucidating the material properties induced by the defects and, for gas sensor and catalytic support applications, spoil their specificity. Thus, better controlled inclusion of defects is necessary to create reproducible defective graphene films.

One-step growth methods based on CVD, over a copper substrate, are the most common approach to growing ideal graphene[16, 17] and recent work by Amontree et al.[18] reinforced the importance of performing graphene CVD growth in an oxygen-free environment, such as ultra-high vacuum (UHV), to obtain high quality and reproducible film growth. While few CVD studies have focused on deliberate inclusion of defects,[19, 20] the choice of precursor is known to have a significant effect on the growth temperature needed to create ideal graphene,[1, 21, 22] thus precursor modification is a promising route for deliberate inclusion of defects. Azupyrene (Figure 1c, inset) can be considered a molecular analogue of the Stone-Wales defect and, were its topology retained in the CVD process, topological defects would be induced into the resulting graphene-like film.

We demonstrate that CVD growth on copper substrates with an azupyrene precursor indeed results in high concentrations of 5-/7-defect sites in a graphene-like film. The defect type is present with a high homogeneity, with vanishingly few alternate defect species present. We obtain continuous monolayer films that can be successfully transferred from their growth substrate onto other supports. These films are obtained without contamination by heteroatoms and at comparatively mild substrate temperatures (~900 K).

**Experiment**

The defective graphene samples were grown in-situ under UHV conditions in the low $10^{-10}$ mbar pressure range. The Cu(111) surface was prepared by sputtering (V = 1 keV, p = 1-5×$10^{-5}$ mbar Ar) and annealing (T = 1000 K). The cleanliness of the crystal was assessed by X-ray photoelectron spectroscopy (XPS), low energy electron diffraction (LEED), and scanning tunnelling microscopy (STM) depending on the system. During growth, the Cu(111) substrate was held at different temperatures and exposed to a molecular flux of the precursor. The precursor azupyrene (synthesis procedure see Ref. [34]) was degassed thoroughly under high vacuum to remove volatile impurities prior to the experiments. It was deposited via a home-built line-of-sight capillary doser (described in Ref. [22]), with the end of the capillary positioned as close to the sample surface as possible under the chamber geometry (within a few millimetres). During deposition, azupyrene was held at a temperature of 25 to 60 °C depending on the deposition geometry and to enable comparable deposition time and coverage. Further experimental and computational details can be found in the supporting information (SI), §1.

## Results and Discussion

## CVD growth of defective carbon networks

Carbonaceous networks were synthesised by CVD of the aromatic precursor azupyrene onto a hot Cu(111) surface. At room temperature, azupyrene adsorbs molecularly on the surface.[8] The adsorption structure and electronic properties of the molecular overlayer have previously been discussed in Refs. [8, 23]. At highly elevated temperatures, ~1000 K, as previously shown,[22, 24] azupyrene forms ideal graphene that demonstrate moiré superstructures, as observed by scanning tunnelling microscopy (STM, Figure 1c). Formation of this film must involve surface-catalysed dehydrogenation and Stone-Wales rearrangement of the 5-/7-membered rings to form a network of 6-membered rings.

At intermediate substrate temperatures, dehydrogenative coupling of azupyrene to form on-surface networks is observed. The film morphology, measured by STM, (Figure 1) varies as a function of growth temperature. Between 700 and 850 K, a dendritic network is formed, imaged (Figure 1a) as bright finger-like protrusions separated by dark regions of the underlying substrate. Bond-resolved, CO-functionalised[25] non-contact atomic force microscopy (nc-AFM) measurements (Figure 1a, inset) show that the 5-/7-topology of azupyrene is retained in these dendrites. Pentagons and heptagons will not tessellate on a plane, thus without significant Stone-Wales re-arrangement an open dendritic network would be expected.

At higher temperatures (~850-950 K) highly corrugated, closed two-dimensional islands are formed (Figure 1b). To tessellate closed films, a large proportion of 6-membered rings must form on the surface and, indeed, the nc-AFM measurements

(Figure 1b, inset) show 5-/7-membered ring defects embedded into a lattice of 6-membered rings. Herein, these closed films are referred to as defective films.

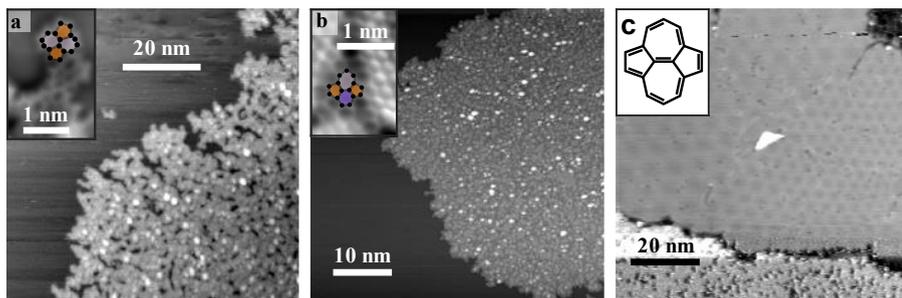

**Figure 1.** Microscopy measurements of the morphological differences in the carbonaceous layers as a function of the growth temperature. STM measurements of (**a**) a dendritic film, (**b**) a defective film and (**c**) ideal graphene. Insets in (**a**) and (**b**) show atomically resolved nc-AFM measurements of (**a**) a dendritic and (**b**) a defective film, highlighting 5-membered (dark orange) rings, and 7-membered (light purple) rings and 6-membered (dark purple) rings. The inset in (**c**) shows the molecular structure of azupyrene. [STM parameters: (**a**) $U_{tip}$ = +1.5 V, $I_{tunnel}$ = 1.25 nA; (**b**) $U_{tip}$ = -1.9 V, $I_{tunnel}$ = 0.5 nA; (**c**) $U_{tip}$ = -1.5 V, $I_{tunnel}$ = 0.8 nA]

## Local defect topology in networks

Large area nc-AFM measurements for a dendritic (Figure 2a-c) and a defective film (Figure 2d-f) indicate that over 50% and 15%, respectively, of all rings are 5-/7-membered. [NB. linking two azupyrene motifs can result in either a 6- or 5-membered ring (SI, §2)]. While different types of 5/7 defects are observed (i.e. other than Stone-Wales), defects with a differing topology (e.g. 4- or 8-membered) are vanishingly rare (<1% of all observed rings) in both films. Enumerating the 4- to 8-membered rings in a dendritic film measured over 59.6 nm$^2$ (see Figure 2a-c, Figure S2 in the SI) yields a distribution of 46 ± 4% 6-membered rings, 32 ± 6% 5-membered rings, 21 ± 6% 7-membered rings, and 1 ± 6% 4- or 8-membered rings (total sample size, N, 283). Similarly enumerating the defective film over 107.5 nm$^2$ (N = 999, see Figure 2d-f, Figures S3-S5 in the SI) yields a distribution of 84 ± 3% 6-membered rings, 9 ± 3% 5-membered rings, and 7 ± 3% 7-membered rings. No vacancy defects were observed. These values are summarised in Table S1 in the SI.

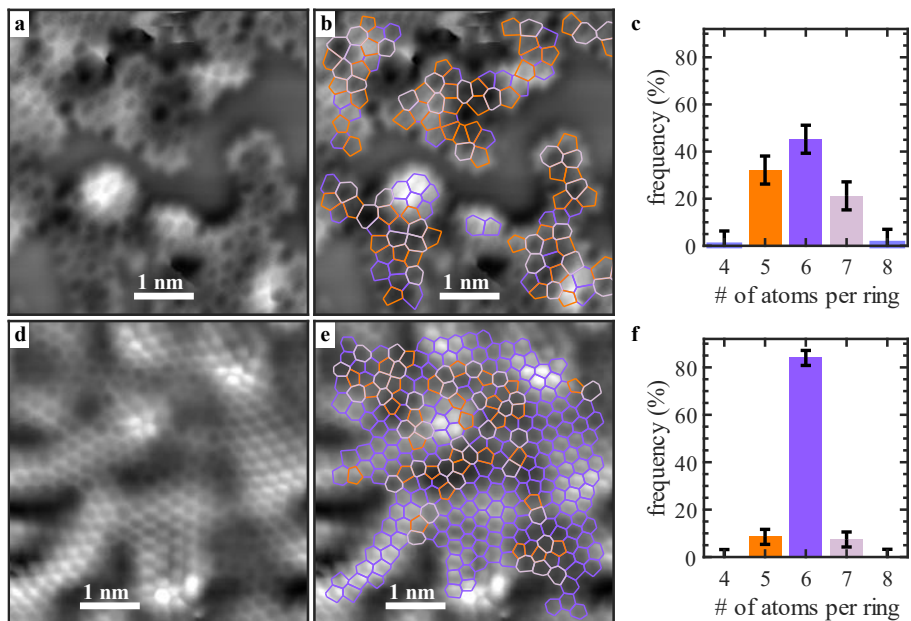

**Figure 2.** Large scale non-contact atomic force microscopy (nc-AFM) demonstrating the defect homogeneity of the topological defects in dendritic and defective films. nc-AFM measurements of (**a**,**b**) the dendritic film and the (**d**,**e**) defective film. Overlayed in panels (**b**) and (**e**) are the C-C bonds, 5-membered rings are dark orange, 7-membered are light purple and 6-membered are dark purple. Histograms of show the relative frequency of 4- to 8-membered rings for (**c**) dendritic and (**f**) defective films. Imaging conditions for the nc-AFM measurements are in the Methods.

## Spectroscopic signal of defective films

Spectroscopic measurements, supported by simulated spectra, show signatures associated with the presence of 5-/7-defects. Soft X-ray photoelectron spectroscopy (SXPS, Figure 3a) indicates a correlation between the binding energy (BE) of the primary peak in the C 1s SXP spectra and the growth temperature and, thus, the concentration of 6-membered rings in the CVD-grown film. The C 1s SXP spectra of molecular azupyrene and graphene adsorbed on Cu(111) (Figure 3a) are consistent with reports in the literature[8, 22, 23, 26] (further discussion in the SI, §3). The C 1s SXP spectrum of the dendritic phase is broader than that of the molecular phase, with a small shoulder at lower BE. The whole spectrum has a BE lower than both graphene and azupyrene. On the high BE tail, the spectrum has a noticeable asymmetry, similar to, though slightly smaller than, that of molecular azupyrene. This asymmetry indicates that an electronic hybridisation with the substrate is, to some degree, present in the dendritic films (see SI, §2). The C 1s SXP spectrum of the defective film has a BE between that of the dendritic phase and ideal graphene. The main peak is narrower than both the dendritic and molecular phase spectra. The shoulder at lower BEs is still present, but is attenuated compared to the dendritic phase. The asymmetry in the high BE tail is not present for the defective films, which may indicate an electronic decoupling of the defective film from the substrate compared to the dendritic phase. Excluding a small transient Sn or Sb contamination, no heteroatoms were observed in the SXPS and the growth is self-limited to a monolayer at high coverage (~40 C atoms/nm$^2$), see SI, §4.

The C K-edge NEXAFS spectra of molecular azupyrene and graphene (Figure 3d) are in agreement with literature data (further discussion in the SI, §5).[8] The C K-edge NEXAFS of the dendritic phase (Figure 3d,f,g) are similar to that of molecular azupyrene, but with broadened features throughout and, in the $\pi^*$ region, an increased absorption rate at the photon energy of the graphene $\pi^*$ peak (blue arrow, Figure 3f). In the $\sigma^*$ region of the spectrum, the dendritic phase has a feature at a lower photon energy than observed for either the molecular phase or ideal graphene (red arrow, Figure 3g). The spectrum of the defective phase is more comparable to graphene, specifically $\pi^*$ features are more pronounced (blue arrow, Figure 3f). While the onset of the $\pi^*$ region is, like the molecular and dendritic phase, at a lower photon energy than ideal graphene (red arrow, Figure 3f), the intensity of this feature is reduced. In the $\sigma^*$ region, the defective film has a sharp feature similar to graphene, but the lower photon energy feature (red arrow, Figure 3g), that was observed in the dendritic phase, is also present.

Simulated XPS data (density functional theory (DFT) calculations, computational details in SI §6 and §7) of a Stone-Wales defect embedded in a graphene lattice qualitatively agree well with the experimentally measured defective SXPS (Figure 3b). These simulations show that the XPS signals attributed to the defect atoms exhibit strong positive and negative BE shifts (Figure 3b,c), with increasingly smaller shifts for atoms that are further from the defect. The 5-membered ring C atoms are shifted to lower BEs, while 7-membered ring C atoms are weakly shifted to higher BEs, interpreted as the rings being electron rich or poor, respectively.[7] The BE shift of the 7-membered ring is larger in a free-standing Stone-Wales defect (Figure S11b in SI).

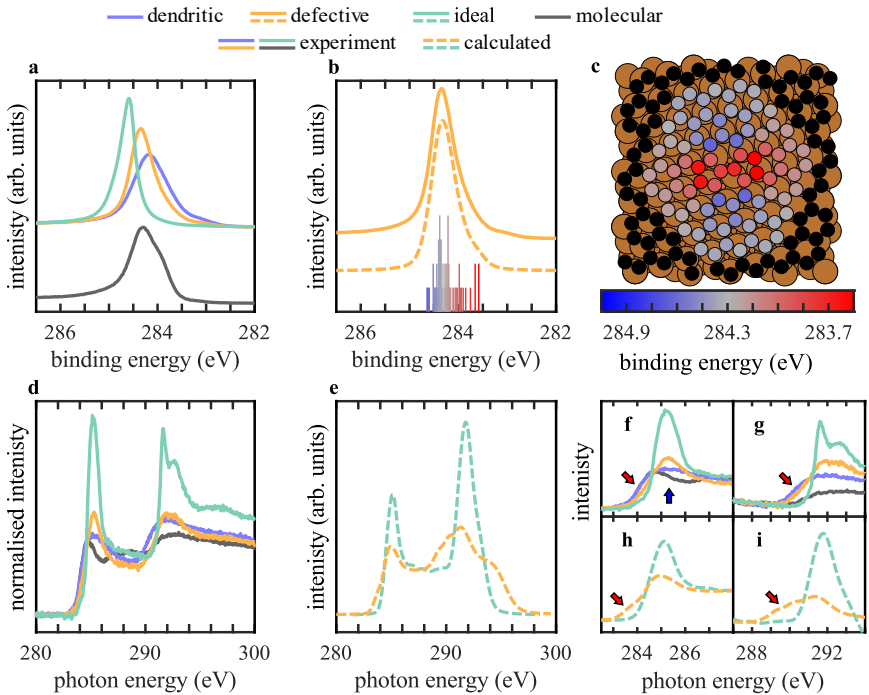

**Figure 3.** Spectroscopic data for graphene-like networks grown on Cu(111) compared to DFT simulation

results. (**a**) Experimental SXPS data, (**b**) comparison of experimental and calculated XPS for defective graphene, with a histogram of the calculated binding energies, without broadening, using the same colour scale as shown in panel (**c**) and (**c**) top view depiction of the structure employed in the DFT calculations, showing the binding energy of each calculated atom as per the indicated colour scale. (**d**) Experimental and (**e**) DFT-simulated C K-edge NEXAFS spectra. Zoomed-in panels of the (**f,h**) $\pi^*$ and (**g,i**) $\sigma^*$ regions for the (**f,g**) experimental and (**h,i**) calculated C K-edge NEXAFS spectra. Further details are in SI, §9. Arrows indicate the regions of the spectra discussed in the main text.

Simulated NEXAFS data (DFT calculations, computational details in SI, §6 and §8) of carbon atoms within the Stone-Wales defect, compared against those within ideal graphene, on Cu(111), also show good qualitative agreement with experiment (Figure 3d-i). As in the experimental data, these simulated spectra indicate that the Stone-Wales defect results in both a broadening of the spectral features, as well as the presence of features at lower photon energies than is observed for ideal graphene. These effects are observed in both the $\pi^*$ and $\sigma^*$ regions (red arrows, Figure 3h,i, respectively), but, in particular, the feature at lower photon energies in the $\sigma^*$ region is only found experimentally for the dendritic and defective phases (red arrow, Figure 3g). This feature is predicted by the theoretical calculations to be associated with the Stone-Wales defect, and may be indicative of a spectral fingerprint for 5-/7-defects.

## Quantitative structure of carbon films

The vertical adsorption structure of the grown films was probed by normal incidence X-ray standing waves (NIXSW, details in SI, §10),[27] Analysis of NIXSW yields two structural parameters, the coherent fraction ($f_H$) and the coherent position ($p_H$). Broadly, $f_H$ is an order parameter, $f_H > \sim 0.7$ indicates a single adsorption height, $f_H < \sim 0.7$ indicates multiple adsorption heights.[28] The $p_H$ is related to the mean adsorption height ($h_H$) by equation (1) in SI, §10.

The NIXSW of graphene and azupyrene adsorbed on Cu(111) are discussed in detail in prior work,[23, 24] but are compared here with respect to exemplar NIXSW data for the dendritic and defective phases (Fig. 4a) measured from the same samples as the STM images in Fig. 1. The NIXSW data show clear and stark differences, related to the different structures present on the surface in each phase.

The resulting mean adsorption heights and coherent fractions of nineteen different samples are shown in Figure 4b,c. Excluding molecular azupyrene, $h_H$ increases roughly linearly with growth temperature. The lowest (highest) adsorption heights, corresponding to the lowest (highest) growth temperatures, are similar to that of molecular azupyrene (ideal graphene, respectively). This variation suggests that the interaction between the substrate and the film decreases as the relative coverage of 5-/7-membered rings decreases. For all defective and dendritic phases, $f_H$ is low ($f_H$=0.2-0.6), with its minimum near the dendritic to defective transition, indicating multiple adsorption heights.[28] Notably, the adsorption height of graphene and azupyrene differ by approximately half of the (111) layer spacing of copper (2.0871 Å). As discussed in SI, §10, an equal mixture of such adsorption heights would result in a $f_H$ of 0. Thus, roughly, the variation in $f_H$ with increasing growth temperature can be understood as the fraction of atoms at an $h_H$ of molecular azupyrene decreasing in favour of those at the $h_H$ of graphene. Illustrating the measured $p_H$ and $f_H$ of the dendritic, defective and ideal graphene films with a simple sinusoidal curve, yields the corrugations displayed in Figure 4d-f.

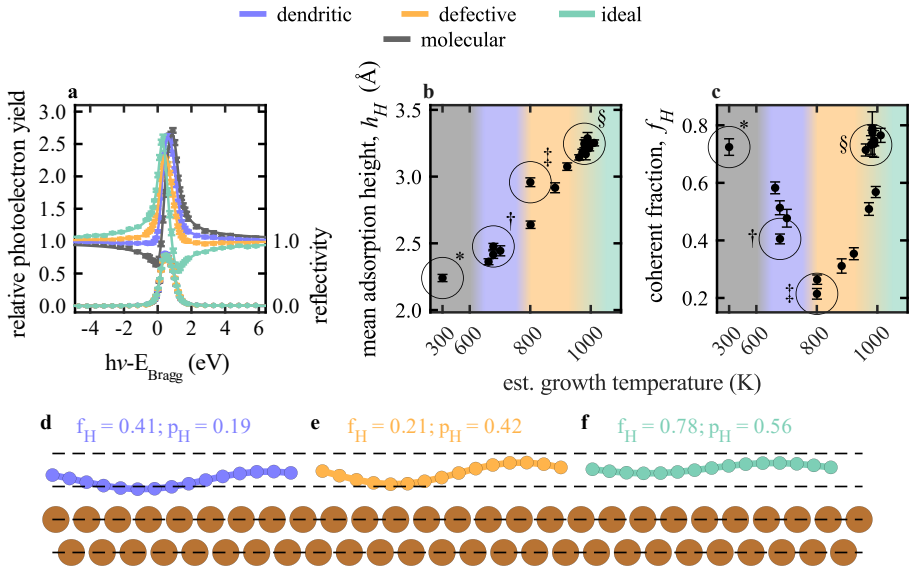

**Figure 4.** Structural data for graphene-like networks grown on Cu(111) at varying temperature. (**a**) Exemplary normal incidence X-ray standing wave (NIXSW) yield curves, dependence of (**b**) mean adsorption height ($h_H$) and (**c**) coherent fraction ($f_H$) on the growth temperature. Data marked in panels (**b,c**) with * (molecular), † (dendritic), ‡ (defective) or § (graphene) correspond, respectively, to the XSW data displayed in (**a**) and were acquired from the same samples as the SXPS data shown in Figure 3a and STM data in Figure 1a-c. The temperature calibration is described in SI, §11. The XPS and XSW data corresponding to panels b and c are shown in Figures S13-19 in the SI. Also shown is an illustration of the measured $f_H$ and $p_H$ for the (**d**) dendritic, (**e**) defective and (**f**) graphene phases with a simple sinusoidal. The $d_{111}$ spacing is indicated with dashed lines.

## Generating freestanding defective films

For many applications of defective graphene-like films, the presence of a copper growth substrate is deleterious and the use of single crystal growth substrates is undesired. Thus, a defective film was grown on copper foil and transferred onto a silicon nitride TEM grid (based on Ref. [29]). Annular dark-field scanning transmission electron microscopy (ADF-STEM) measurements over 1406 nm² (N = 6505, Figure 5, Figures S20-24 in the SI) demonstrate that the 5-/7-membered topological defects are present (11 ± 1% and 10 ± 1%, respectively – details in the SI, §12). These results confirm that this CVD method can grow the same quality of films on Cu foils, which can then be transferred onto a wide array of substrates for different technological applications.

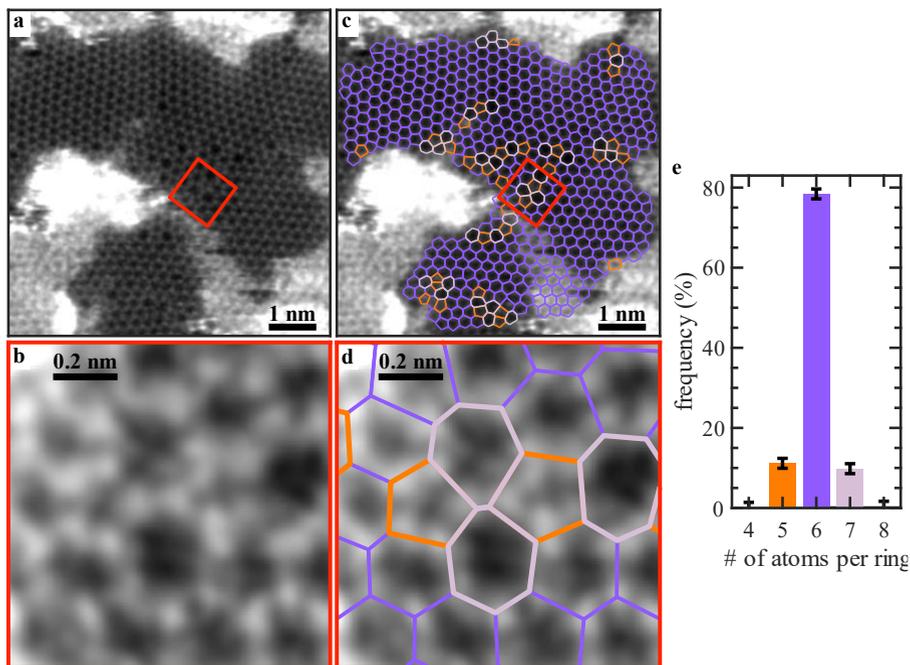

**Figure 5.** Annular dark-field scanning tunnelling electron microscopy (ADF-STEM) showing the retention of 5-/7-topological defects after transfer onto a TEM grid. **(a-d)** Atomically resolved ADF-STEM measurements of a defective film. Overlayed in panels **(c)** and **(d)** are the C-C bonds, following the same colour legend as Figure 2. A single Stone-Wales defect is highlighted in panels **(b)** and **(d)** and corresponds to the red box indicated in **(a,c)**. **(e)** Histogram showing the relative frequency of 4- to 8-membered rings.

# Discussion and Conclusions

We have demonstrated that topological defects in a graphene-like film can be grown with a high homogeneity of defect type by using a molecular precursor, azupyrene, that shares the topology of that defect, specifically 5-/7-membered carbon rings. Between ~700-850 K a dendritic network is formed; between ~850-950 K a closed film with a defect abundance of ~20% are formed. Unlike prior work,[17, 19] these films are grown without potential halide contaminants and form self-limited monolayers at high coverage.

Unlike post-processing of graphene oxide[30, 31] or ideal graphene,[11, 12, 31, 32] which can result in significant defect inhomogeneity (e.g. topological, vacancy, heteroatoms and non-aromatic C defects) and Ullmann coupling methods[13, 14, 15] that can result in significant halide contamination, the one-step CVD method with topological precursors, presented here, yields films with unprecedentedly high defect homogeneity. To the author's knowledge, the same scale of defect homogeneity, as observed in this study, has not been reported previously.

While the importance of defects in graphene is understood for several applications,[2, 4, 5] understanding which defects play what role is an important unanswered question. There is clear need for samples with a high defect type homogeneity that can be transferred onto technologically relevant substrates, as demonstrated here. We posit that this core methodology, of growing defective graphene films utilising a molecular precursor that shares a moiety with the defect, can be generalised to other defect types. For example: using a heteroatomic precursor could generate heteroatomic doping of the graphene film, as has been observed with nitrogen and boron containing molecules[20, 33]; using a precursor with a C vacancy template, could generate graphene with a high number of C vacancy sites. Potentially, designing defects into graphene layers may be as simple as synthesising molecular precursors with that signature moiety. Close-to defect-pure graphene samples can crucially support mechanistic studies in catalysis and gas sensing to optimise binding specificity and catalytic selectivity. The presented growth approach has the potential to deliver scalable production of highly functional purpose-designed materials.

**Supporting information.** The supporting information is available at XXX free of charge and contains additional methodological details on the sample preparation, experimental measurements and their analysis and the DFT calculations, as well as further nc-AFM, ADF-STEM and NIXSW results that are summarised in this article.

**Acknowledgments.** The authors thank Diamond Light Source for access to beamline I09 (SI25379-4, SI30875-1, SI31695-3 NT31165-2 and NT33709-1) and for access and support in the use of the electron Physical Science Imaging centre (E02, proposal number NT39165). We also thank MAX IV synchrotron for access to the beam line FlexPES (20200269). The authors thank Stephan Appelfeller for their support during the beamtime at Max IV. The authors acknowledge use of the Raman spectrometer as well as the support by Ben Breeze of the Spectroscopy Research Technology Platform, University of Warwick. We also thank Raymundo Marcial Hernandez and Christian Bech Nielsen of Queen Mary University London for fruitful discussions and their support in this work.

Financial support is acknowledged from the DFG under the Walter Benjamin fellowship programme [KL 3430/1-1] (B. P. K.), Analytical Science CDT at the University of Warwick and the Diamond Studentship program (R. J. M. and D. A. D.), the EPSRC-funded CDT for Modelling of Heterogeneous Systems (HetSys CDT) [EP/S022848/1], the UKRI Future Leaders Fellowship [MR/S016023/1, MR/X023109/1] and a UKRI Frontier grant [EP/X014088/1] (R. J. M.), a New Investigator Award from the Engineering and Physical Sciences Research Council [EP/X012883/1] (D. A. D. and F. E.), and a Royal Society University Research Fellowship (A. S.).

High-performance computing resources were provided via the Scientific Computing Research Technology Platform of the University of Warwick and the following EPSRC-funded UK High-End Computing Consortia: Materials Chemistry Consortium [EP/R029431/1, EP/X035859/1] and the HPC-CONEXS consortium [EP/X035514/1].

## Author Contributions

The STM data were acquired by B. P. K, M. A. S, L. A. R. and D. A. D.; analysed and interpreted by B. P. K and D. A. D; under supervision by D. A. D. The nc-AFM data were acquired, analysed and interpreted by B. P. K., M. A. S. and J. D.; under supervision by W. A.. The XPS and NIXSW data were acquired by B. P. K., M. A. S., L. A. R., T. L., L. B. S. W. and D. A. D.; analysed and interpreted by B. P. K. and D. A. D.; under supervision by T-. L. L. and D. A. D.. The NEXAFS data were acquired by B. P. K., M. A. S., A. P. and D. A. D.; analysed and interpreted by B. P. K. and D. A. D.; under supervision by D. A. D.; A. G. created new software for the acquisition of the NEXAFS data. The TEM data were acquired by D. H. and C. A.; analysed by F. E. and D. A. D., with support from A. S.; and interpreted by D. A. D.. The samples were grown by B. P. K., M. A. S., L. A. R., TL, L. B. S. W. and D. A. D.; under supervision by D. A. D.. The azupyrene was synthesised by L. S. and S. M. W.; under supervision by G. H.. The samples were transferred to TEM grids by S. S. A.; under supervision by R. G. and S. J. H.. DFT calculations were performed by B. P. K., M. A. S. and D. B. M.; interpreted by B. P. K. and R. J. M.; under supervision by R. J. M.. The study was conceived and designed by B. P. K., R. J. M. and D. A. D.. The paper was primarily drafted by D. A. D., with significant support from B. P. K., D. B. M. and R. J. M.. The paper was revised by all authors.

## Declarations

The authors have no conflicts to disclose.

# One-step synthesis of graphene containing topological defects

Benedikt P. Klein[1,2†], Matthew A. Stoodley[1,2], Joel Deyerling[3], Luke A. Rochford[1,4], Dylan B. Morgan[2], David Hopkinson[1], Sam Sullivan-Allsop[5], Fulden Eratam[1], Lars Sattler[6], Sebastian M. Weber[6], Gerhard Hilt[6], Alexander Generalov[7], Alexei Preobrajenski[7], Thomas Liddy,[1,8] Leon B. S. Williams[1,9,10], Tien-Lin Lee[1], Alex Saywell[11], Roman Gorbachev[5], Sarah J. Haigh[5], Christopher Allen[1,12], Willi Auwärter[3], Reinhard J. Maurer[2,13*] and David A. Duncan[1*‡]

[1]Diamond Light Source, Harwell Science and Innovation Campus, Didcot, OX11 0DE, United Kingdom.
[2]Department of Chemistry, University of Warwick, Gibbet Hill Road, Coventry, CV4 7AL, United Kingdom.
[3]Physics Department E20, TUM School of Natural Sciences, Technical University of Munich, James-Franck-Straße 1, 85748 Garching, Germany.
[4]Department of Earth Sciences, University of Cambridge, Downing Street, Cambridge, CB2 3EQ, United Kingdom.
[5]National Graphene Institute, University of Manchester, Oxford Road, Manchester, M13 9PL, United Kingdom.
[6]Institute of Chemistry, Carl von Ossietzky University Oldenburg, Carl-von-Ossietzky-Straße 9-11, 26111 Oldenburg, Germany.
[7]MAX IV Laboratory, University of Lund, Fotongatan 2, 224 84 Lund, Sweden
[8]School of Chemistry, University of Nottingham, University Park, Nottingham, NG7 2RD, United Kingdom
[9]School of Chemistry, University of Glasgow, University Avenue, Glasgow, G12 8QQ, United Kingdom.
[10]School of Physics & Astronomy, University of Glasgow, University Avenue, Glasgow, G12 8QQ, United Kingdom.
[11]School of Physics & Astronomy, University of Nottingham, University Park, Nottingham, NG7 2RD, United Kingdom
[12]Department of Materials, University of Oxford, Parks Road, Oxford, OX1 3PH, United Kingdom.
[13]Department of Physics, University of Warwick, Gibbet Hill Road, Coventry, CV4 7AL, United Kingdom.
*Corresponding authors. E-mails: r.maurer@warwick.ac.uk; david.duncan@nottingham.ac.uk;
†Current address: Research Center for Materials Analysis, Korea Basic Science Institute, 169-148 Gwahak-ro, Yuseong-gu, Daejeon 34133, Republic of Korea
‡ Current address: School of Chemistry, University of Nottingham, University Park, Nottingham NG7 2RD, United Kingdom

# Supporting Information

**Table of Contents**

## 1. Methods
### 1.1 STM measurements
STM measurements were performed at the Surface Interface Laboratory (SIL) at the Diamond Light Source (DLS) using an Omicron VT STM with an etched tungsten tip at room temperature. The base pressure in the chamber was $5 \times 10^{-10}$ mbar and samples were prepared in situ in the same vacuum system. STM measurements were performed using an Omicron VT STM

## 1.2 SXPS measurements

SXPS measurements were performed at the I09 beamline at the synchrotron Diamond Light Source.[1] Samples were either prepared in situ in the same vacuum system or transferred from the SIL preparation chamber via vacuum transfer. The SXPS data was recorded using the synchrotron light utilising a Scienta EW4000 HAXPES hemispherical electron energy analyser mounted perpendicular to the incident direction of the X-ray radiation in the same plane as the photon polarisation (linear horizontal). A photon energy of 430 eV was used for the C 1s spectra, the binding energy was corrected by performing a Fermi energy (FE) measurement at the same photon energy and pass energy, and setting the FE to be the origin of the binding energy scale.

## 1.3 nc-AFM measurements

nc-AFM measurements were performed at the Physics Department of the Technical University of Munich (TUM) using a commercial 4K LT-STM / nc-AFM Createc microscope, operated at approximately 6 K during measurements. This microscope contains a qPlus sensor[2] (frequency-modulation mode at an oscillation amplitude of 60 pm) for performing the nc-AFM measurements, which were performed in constant high mode (open feedback loop) at sample bias of approximately 0 V. Samples were prepared in situ in the same vacuum system. For reliable / easier tip functionalisation, NaCl islands were grown (sample temperature -10 to 10 °C) after the initial sample preparation with azupyrene. The NaCl grew predominately on graphene-free Cu(111) areas. CO was dosed onto the cold sample (< 10 K) and the tip was functionalised by picking up CO from the NaCl islands. The nc-AFM measurements of the dendritic and defective film were each performed on a single sample over multiple spots on those two samples. The abundance of 5-, 6- and 7- membered rings was determined by Voronoi tessellation. The centre of the ring, used as the Voronoi seed, was identified manually, areas of the images that were deemed too challenging to identify these positions were masked (e.g. the region circled in red in Figure S5 in the SI) and the resulting Voronoi diagram was composed of the unmasked regions. The MATLAB built in voronoi command, inputted with the identified position of the ring centres in the image, was used to create the Voronoi diagram and provided the vertices of the rings and thus its associated topology. The counting of the data was assumed to be Poisson distributed and thus the uncertainty in the enumerated number of rings of each topology was assumed to be the square root of the number of counted rings.

## 1.4 NEXAFS measurements

NEXAFS measurements were recorded in partial electron yield (PEY) mode at the FlexPES beamline at the synchrotron MAX IV.[3] The PEY detector was a multi-channel plate (MCP) mounted below the sample. A retardation grid was mounted on the MCP to act as a high-pass filter, rejecting low energy electrons, resulting in improved surface sensitivity.[4] For the measurements presented here a retardation bias of 150 V was used. Each spectrum was measured between 5-10 times across different spots on the sample. These data were integrated together and then normalised by the drain current from a clean gold monitor, mounted between the photon source and the sample. A comparable spectrum was measured from the clean sample and this "crystal spectrum" was then normalised to have the same intensity in the pre-edge region 275.0-282.5 eV as the experimental spectrum. The crystal spectrum was then subtracted from the experimental spectrum and the resulting spectrum was normalised to the average intensity over the photon energy range of 320-325 eV.

## 1.5 NIXSW measurements

NIXSW measurements were performed at the I09 beamline at the synchrotron Diamond Light Source. Samples were either prepared *in situ* in the same vacuum system or transferred from the SIL preparation chamber via vacuum transfer. The measurement setup for the photoelectron yield was the same as for the regular XPS data, due to the requirements of NIXSW, the hard X-ray branch of I09 was used. The reflectivity was acquired simultaneously to the photoelectron yield using a fluorescent plate that is mounted inside the flange through which the incident photon pass. The measured reflectivity curve was used to define the position of the Bragg energy, as well as the broadening present in the system due to imperfections in the monochromator or the Cu(111) single crystal. The non-dipolar effects in the NIXSW measurement[5] were addressed using a so-called "backwards-forwards Q-parameter",[6] which was calculated theoretically[7] using the angle between photon polarisation and the median photoelectron intensity emission angle ($\theta = 18°$). The individual energy distribution curves of the NIXSW measurements were fitted with a convolution of a Gaussian and a Doniach-Sunjic line shape[8]; and a Gaussian error function, that shared the same Gaussian width as the peak, was used to model the step in the photoelectron intensity before and after the peak. A straight line was used to model the background variation. The area of the fitted peak was then used as the photoelectron yield and plotted relative to their off-Bragg photoelectron yield.

## 1.6 Experiments on polycrystalline Cu foils and ADF-STEM measurements

The defective graphene sample on polycrystalline Cu foil (Advent Research Materials, 99.996+% purity, 0.025 mm thick) was grown in-situ in the sample preparation chamber of the I09 beamline with pressure in the low $10^{-10}$ mbar range. The Cu foil was prepared by repeated sputter (V = 1 keV, p = $2 \times 10^{-5}$ mbar Ar) and annealing (T = 1000 K) cycles. The cleanliness of the foil was assessed by XPS. After growth (~700 K) and XPS measurements, the foil was removed from vacuum and transported in an inert atmosphere to a clean room at the National Graphene Institute in Manchester. The defective graphene film was transferred to the TEM grid using the below procedure. The copper foils were spin-coated (3000 rpm, 1 min) with PMMA/Anisole solution (8%) sample side up, and subsequently set on a hot plate (130 °C, 5 min). Two pieces of tape, with holes in the centre, were attached to the PMMA coated foil. The foils were submerged in the APS solution for 21 hours to allow complete etching of the copper. Each sample was then transferred to a silicon nitride TEM grid via a transfer rig. Once attached, the TEM grid was then heated to 75 °C to adhere the sample to the grid and the tape removed. The TEM grid was then heated to 130 °C on a hot plate for 5 mins. Finally, the PMMA was removed by being stirred in acetone (50 °C, 5 min) before subsequently being dipped in acetone (RT), warm IPA (50 °C), allowed to dry and then dipped in toluene (RT).

ADF-STEM measurements were performed using a JEOL ARM300CF microscope operating at 80 keV with convergence semi-angle of 25 mrad, ADF collection angular range of 47 to 170 mrad and a beam current of approximately 30 pA.

The images were filtered by 2D-Gaussian filtering (MATLAB 2023a, imgausfilt) whose two dimensional smoothing kernel had a standard deviation of 3.5 pixels. The abundance of 5-, 6- and 7- membered rings was determined by Voronoi tessellation using the same procedure as detailed in the nc-AFM section.

## 1.7 Density Functional Theory Calculations

Density Functional Theory calculations have been performed with the FHI-aims (v210716)[9] and CASTEP[10] (version 21.1) codes using the PBE exchange-correlation functional[11]. FHI-aims was used for geometry optimizations. Optimized structural models of freestanding ideal graphene, Cu(111)-adsorbed ideal graphene, freestanding Stone-Wales-defective graphene and Cu(111)-adsorbed Stone-Wales defective graphene are based on previous work and built from a Moiré superstructure of a $3\sqrt{13} \times 3\sqrt{13}$ Cu(111) surface slab that accommodates a $2\sqrt{31} \times 2\sqrt{31}$ graphene layer with an angle of rotation of 5°.[12] CASTEP with standard on-the-fly generated ultrasoft pseudopotentials was used to perform XPS and NEXAFS simulations for the optimized structures. XPS simulations were performed with the Delta-Self-Consistent-Field (Δ-CF) method and NEXAFS simulations were performed with the ionisation-potential-corrected transition-potential method (ΔIP-TP).[13, 14] Spectra were broadened with a pseudo-Voigt function to capture instrumental and lifetime broadening effects.[15] For the SXPS modelling, a Gaussian full width half maximum (FWHM) of 0.40 eV and a Lorentzian FWHM of 0.46 eV were used. The NEXAFS spectra were broadened with a pseudo-Voigt function with an energy-dependent broadening model, starting with a 0.75 eV broadening at an 80%/20% Gaussian-Lorentzian ratio at the leading edge. Further details on Density Functional Theory calculations and core-level spectra simulations are given in §5-7.

## 2. Simplistic Homocoupling of azupyrene

In the case of a simple homocoupling of two azupyrene molecules there are three symmetrically unique bonding configurations, as shown in Figure S1. Two configurations result in a 5-membered ring being formed, one results in a 6-membered ring. There are 16 possible combinations of the configurations in Figure S1a that yield a 5-membered ring and the configuration in Figure S1c that yields a 6-membered ring, but only 8 combinations that will yield the configuration in Figure S1b. Thus, by considering this first linking stage, one would naively expect homocoupling to result in 50% more 5-membered rings, than 6-membered, rings. Regardless of the actual ratio of 5- to 6-membered rings found to link these molecules, that 7-membered rings cannot form in this way suggests why in both dendritic and defective films 5-membered rings are found to be more numerous.

**Figure S1**

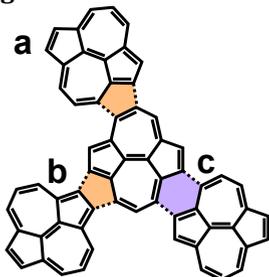

**Figure S1:** Possible bonding configurations of two azupyrene molecules. Solid lines indicate bonds within the azupyrene molecules, hashed lines indicate possible bonds formed by dehydrogenative homocoupling of the molecules. As shown there are two possible configurations that result in (**a,b**) a 5-membered ring and one configuration that results in (**c**) a 6-membered ring. In this simplified two molecule case, there is no possible configuration that yields a 7-membered ring.

**Figure S2**

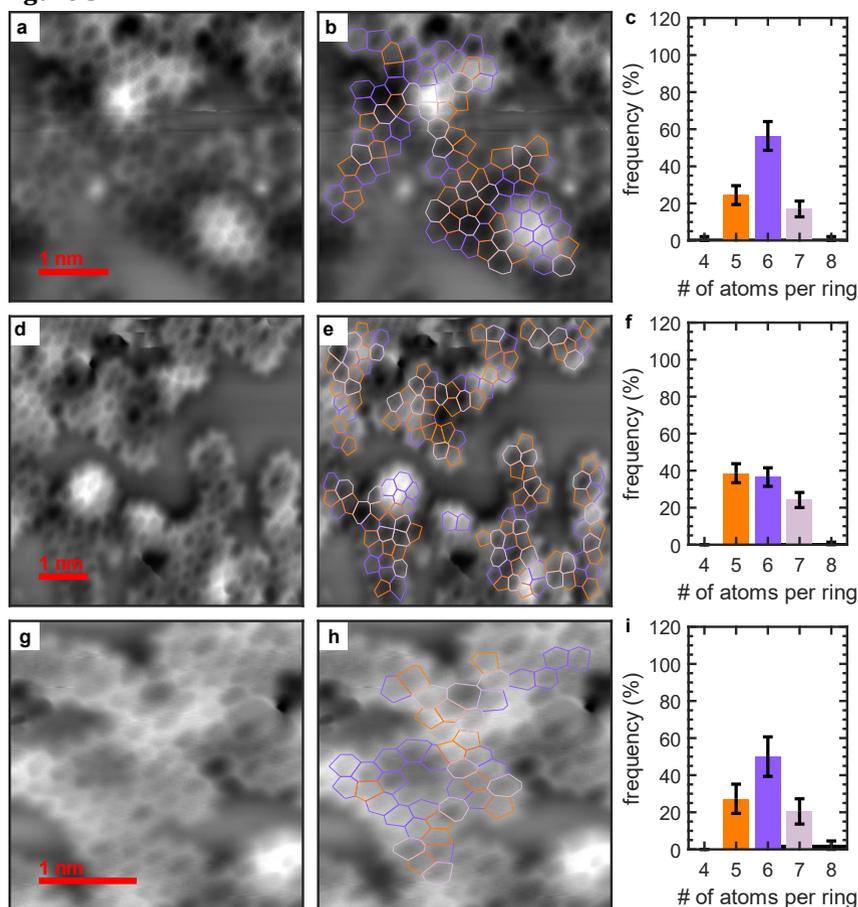

**Figure S2:** (**a-b,d-e,g-h**) Atomically resolved nc-AFM measurements of a dendritic film. Overlayed in panels (**b,e,h**) are the C-C bonds, 5-membered rings are dark orange, 7-membered are light purple and 6-membered are dark purple. Also shown in panels (**c,f,i**) are the histograms of the relative frequency of 4- to 8-membered rings. Imaging conditions are stated in the methods section of the main article.

**Figure S3-5**

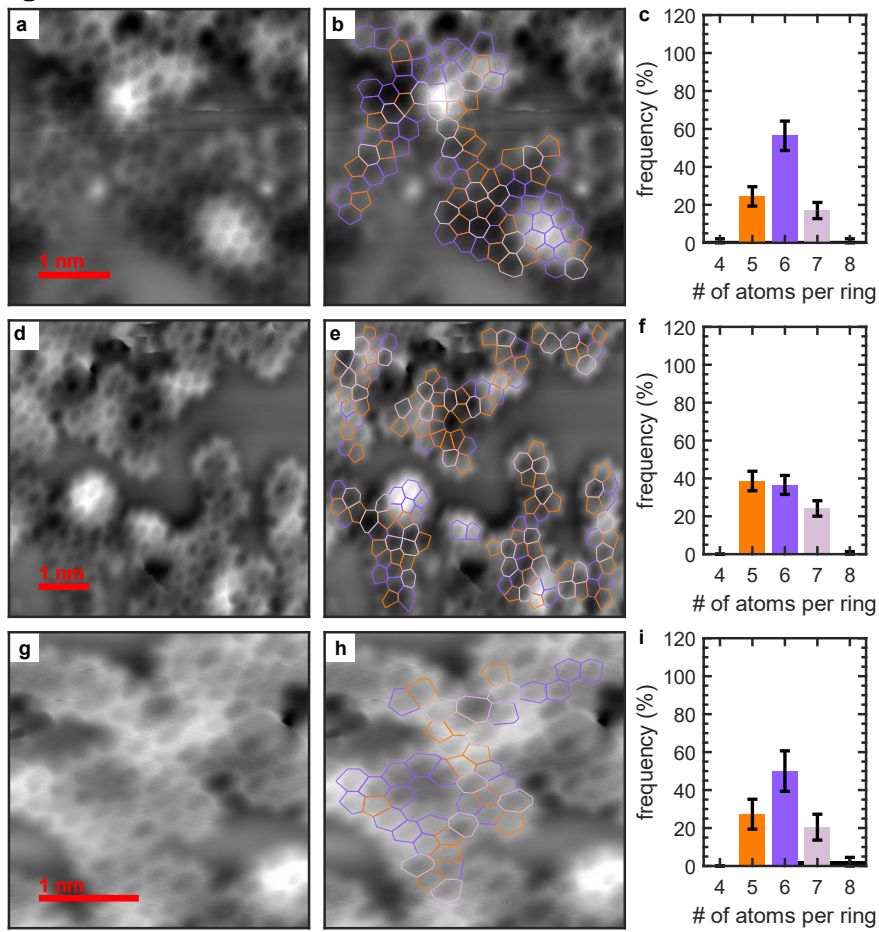

**Figure S3:** (**a-b**,**d-e**,**g-h**) Atomically resolved nc-AFM measurements of a defective film. Overlayed in panels (**b**,**e**,**h**) are the C-C bonds, 5-membered rings are dark orange, 7-membered are light purple and 6-membered are dark purple. Also shown in panels (**c**,**f**,**i**) are the histograms of the relative frequency of 4- to 8-membered rings. Imaging conditions are stated in the methods section of the main article.

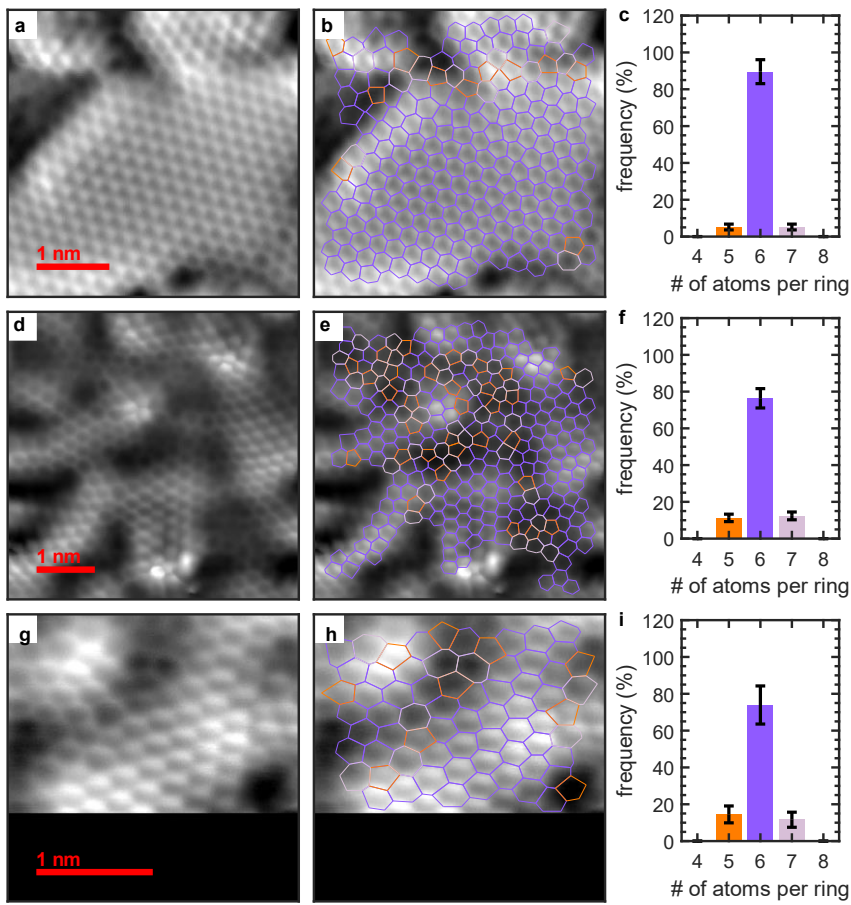

**Figure S4:** (**a-b,d-e,g-h**) Atomically resolved nc-AFM measurements of a defective film. Overlayed in panels (**b,e,h**) are the C-C bonds, 5-membered rings are dark orange, 7-membered are light purple and 6-membered are dark purple. Also shown in panels (**c,f,i**) are the histograms of the relative frequency of 4- to 8-membered rings. Imaging conditions are stated in the methods section of the main article.

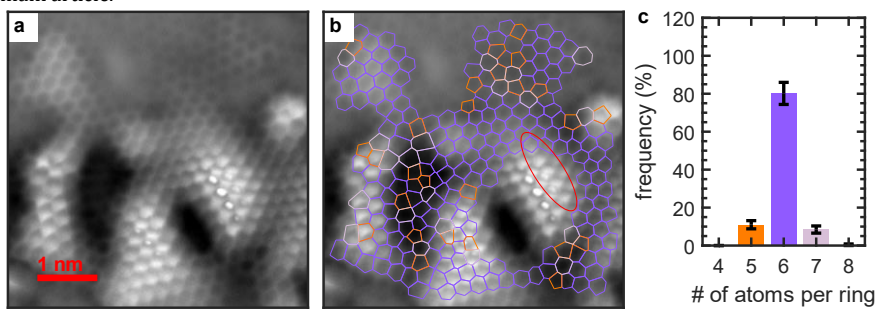

**Figure S5:** (**a,b**) Atomically resolved nc-AFM measurements of a defective film. Overlayed in panel (**b**) are the C-C bonds, 5-membered rings are dark orange, 7-membered are light purple and 6-membered are dark purple. Also shown in panel (**c**) is the histograms of the relative frequency of 4- to 8-membered rings. Imaging conditions are stated in the methods section of the main article. Circled in red are a row of six membered rings that transition from the centre of the ring contrasting darkly, to the centre of the ring contrasting brightly, hindering local assignment of the topology.

**Table S1**

**Table S1:** The enumeration, both in raw number and percentage, of the number of rings observed that are 4- to 8-membered for the defective and dendritic films measured by nc-AFM, as well as the defective film measured by ADF-STEM. The value in brackets is the number of rings identified with that number of atoms / bonds. The error, in percentage, for the nc-AFM: dendritic sample is 6%, nc-AFM : defective is 3% and ADF-STEM : defective is 1%.

| technique | film | 4- | 5- | 6- | 7- | 8- |
|---|---|---|---|---|---|---|
| nc-AFM | dendritic | 0% (1) | 32% (91) | 45% (128) | 21% (60) | 1% (3) |
| nc-AFM | defective | 0% (0) | 9% (85) | 84% (839) | 7% (74) | 0% (1) |
| ADF-STEM | defective | 0% (11) | 11% (726) | 78% (5101) | 10% (639) | 0% (28) |

## 3. Sources of asymmetrical loss features in X-ray photoelectron spectroscopy

The soft X-ray photoelectron spectroscopy (SXPS) obtained from molecular azupyrene and graphene grown on Cu(111) are shown in Figure 3a of the main manuscript. The graphene C 1s SXPS is found at a higher BE than the molecular azupyrene and consists of a single, narrow peak, indicating closely similar chemical environments for each carbon atom. The C 1s SPXS of molecular azupyrene exhibits a shoulder at lower BE, that has previously been discussed in detail.[16] The C 1s SXPS spectra for both, molecular azupyrene and ideal graphene, show asymmetry to higher BEs in the lineshape, which has previously been assigned to shake-up excitation of bound electrons close to the Fermi energy, as discussed below.[17] Graphene on Cu is n-doped,[18] while the molecular orbitals of azupyrene hybridise with the delocalised electronic states of the Cu substrate,[19] resulting in both systems having an electronic density of states around the Fermi energy. This electronic configuration allows gapless excitation of electrons from occupied to unoccupied states, such that the emitted photoelectron can lose kinetic energy to promote these excitations, resulting in an asymmetrical XPS lineshape.

## 4. XPS measurements excluding contaminants and indicating self-limiting growth

In addition to the high resolution C 1s SXPS data shown in the main manuscript, wide range overview spectra were also acquired, shown in Figures S6-S8. In these spectra no first or second row p-block contaminant (e.g. B, N, O, Si, F, Cl, Br or I) was observed. For some samples a small Sn or Sb contaminant (maximum coverage ~1 Sn atom per 100 surface Cu atoms, in comparison to the Cu 3s core level) was observed, due to outgassing of the Mo clips that held the sample in place. This contaminant was transient, i.e. different batches of the Mo clips either did or did not have such a contamination. No systematic differences were observed between samples that contained these trace Sn / Sb contaminants and those that did not. No other contaminants were observed in the as-grown samples. We observed a self-limiting growth for the defective graphene. Figure S9 shows the SXPS, of both a wide range overview spectra (Figure S9a) and a C 1s core level (Figures S9b) after deposition at intermediate temperatures for 90 and 180 minutes. After a 90 minute deposition the coverage is close to that expected for a saturated carbon monolayer (comparing the C 1s to Cu 3p intensity suggests an atomic density of the C atoms of ~40 atoms/nm$^2$. NB. that this estimate is based on the overview spectra shown in Figure S9a and will therefore have a large associated uncertainty). Doubling the growth time to 180 minutes resulted only a very small (< 1%) increase in the area of the C 1s signal. Furthermore, post annealing of the film in vacuum to elevated temperatures where graphene growth, under an azupyrene flux, is observed, resulted in no apparent change to the C 1s signal. This lack of change indicates that the structure of the network is determined by the temperature during growth and that the grown defective films are thermally stable up to elevated temperatures.

**Figures S6-S8**

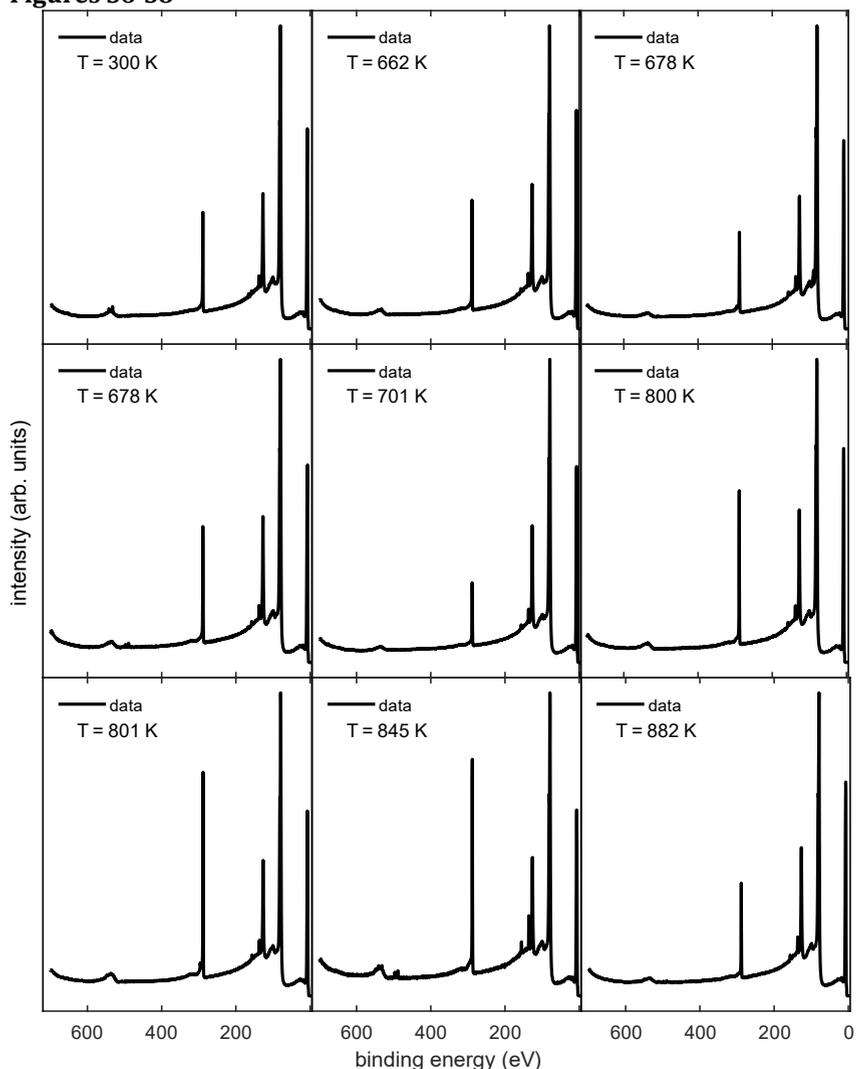

**Figure S6:** Overview SXPS data taken from the samples for which the NIXSW data was performed (see Figure 4b,c in main manuscript). Indicated is the estimated substrate growth temperature for each sample. All spectra were obtained at a photon energy of 850 eV and the binding energy is uncorrected. Sharp features at ~450 eV correspond to Sn contamination, ~550 eV to Sb contamination. The broad feature at ~550 eV is the C KLL Auger decay spectra. Features at 136 and 156 eV binding energy are the Cu 2p spectra originating from the second order light from the soft X-ray monochromator ($h\nu = 1700$ eV).

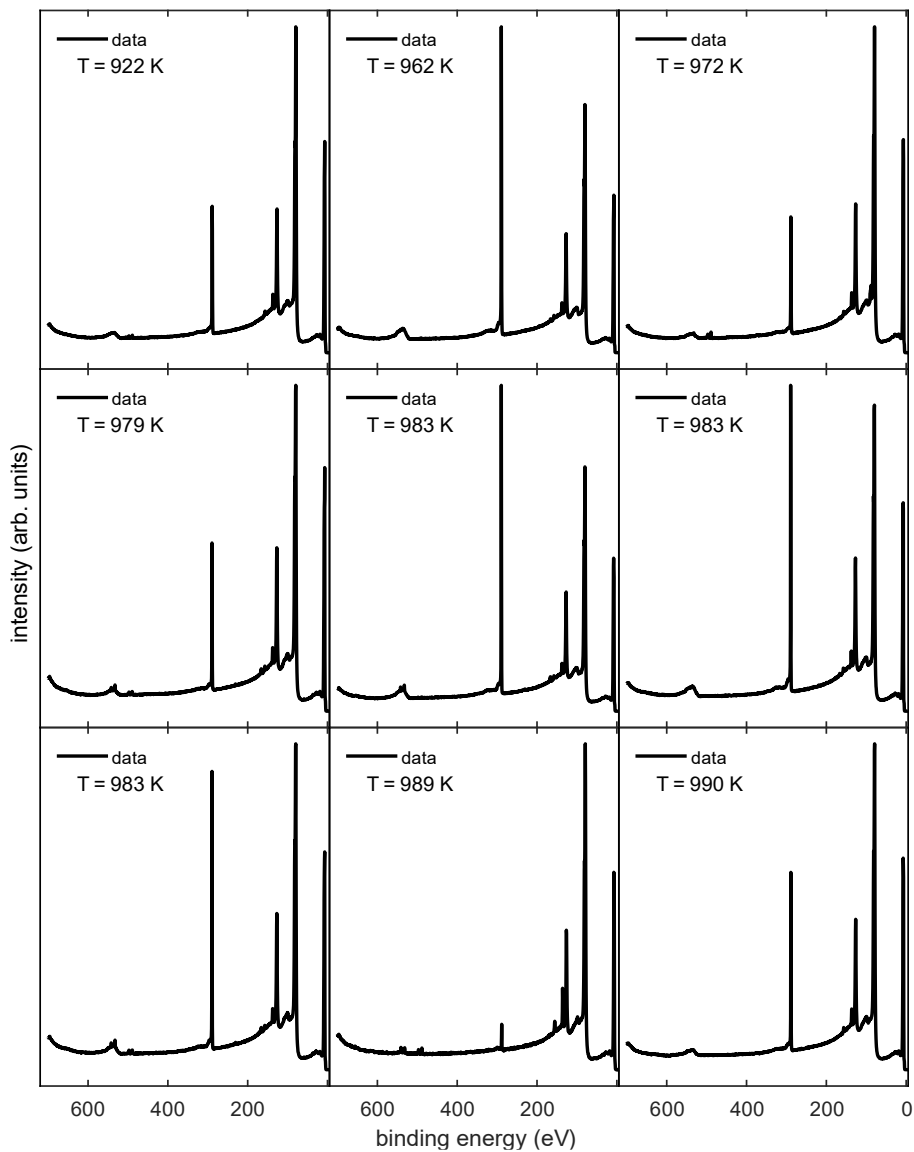

**Figure S7:** Overview SXPS data taken from the samples for which the NIXSW data was performed (see Figure 4b,c in main manuscript). Indicated is the estimated substrate growth temperature for each sample. All spectra were obtained at a photon energy of 850 eV and the binding energy is uncorrected. Sharp features at ~450 eV correspond to Sn contamination, ~550 eV to Sb contamination. The broad feature at ~550 eV is the C KLL Auger decay spectra. Features at 136 and 156 eV binding energy are the Cu 2p spectra originating from the second order light from the soft X-ray monochromator ($hv = 1700$ eV).

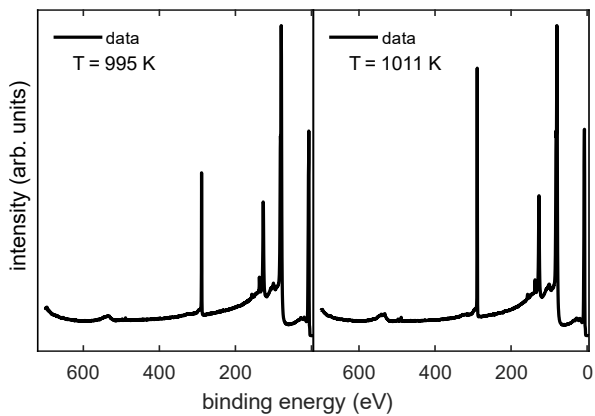

**Figure S8:** Overview SXPS data taken from the samples for which the NIXSW data was performed (see Figure 4b,c in main manuscript). Indicated is the estimated substrate growth temperature for each sample. All spectra were obtained at a photon energy of 850 eV and the binding energy is uncorrected. Sharp features at ~450 eV correspond to Sn contamination, ~550 eV to Sb contamination. The broad feature at ~550 eV is the C KLL Auger decay spectra. Features at 136 and 156 eV binding energy are the Cu 2p spectra originating from the second order light from the soft X-ray monochromator ($h\nu = 1700$ eV).

**Figure S9**

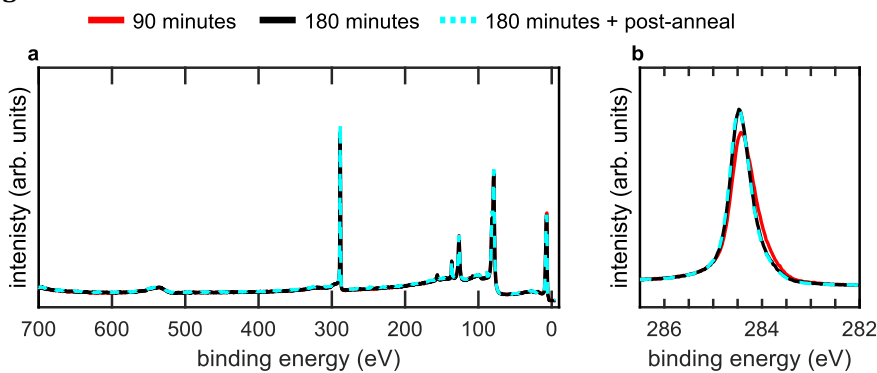

**Figure S9:** (**a**) Overview and (**b**) C 1s SXPS data from defective graphene films grown for protracted deposition times (90, red line, and 180 minutes, black line), indicating that the film is self-limited to a single monolayer at high coverage. Also shown is the result of post annealing (dashed, cyan line) a defective graphene film in vacuum, to temperatures where the growth of graphene under an azupyrene flux is observed, resulting in no observable changes in the spectra.

## 5. Further discussion on the experimental NEXAFS data

The near-edge X-ray absorption fine structure (NEXAFS) spectroscopy obtained from molecular azupyrene and graphene grown on Cu(111) are shown in Figure 3d of the main manuscript. In both the $\pi^*$ region (Figure 3f) and the $\sigma^*$ region (Figure 3g) clear differences are present between molecular azupyrene and graphene. In the $\pi^*$ region, the first peak of molecular azupyrene occurs ~0.7 eV lower in photon energy than graphene and, in the $\sigma^*$ region, instead of a sharp resonant feature, a broad one is observed instead for azupyrene.

## 6. Details on the computed XPS and NEXAFS data: Model structures

Surface slab models of metal-adsorbed graphene were built based on previously moiré superstructure of a $3\sqrt{13} \times 3\sqrt{13}$ Cu(111) surface slab that accommodates a $2\sqrt{31} \times 2\sqrt{31}$ graphene layer with an angle of rotation of 5°.[12] The total number of carbon atoms is 248. The total number of copper atoms is 936. Structures were relaxed with the all-electron code FHI-aims[9] and the PBE+MBD-NL functional[20] with other settings reported in detail in Ref. [12]. Surface slab models were optimized with eight metal layers where the bottom six layers were frozen in their bulk truncated structure. Using this approach, we created relaxed structures for a free-standing ideal graphene layer, a free-standing Stone-Wales defective graphene layer, ideal graphene adsorbed on Cu(111) and a graphene layer with a Stone-Wales defect centred atop a bridge site. The corresponding structures are shown in Figure S10. For the subsequent XPS simulations, the number of metal layers was reduced to four. For the NEXAFS simulations, the number of metal layers was reduced to two.

## 7. Details on the computed XPS and NEXAFS data: XPS calculations

XPS and NEXAFS C 1s spectra were simulated with core hole calculations based on the optimized structures with the CASTEP code, version 21.1[10] using the PBE exchange-correlation functional[11]. In each calculation, a core-hole-excited pseudopotential was created for the target atom and the calculation was repeated for each targeted atom. We closely follow the procedure explained in detail in Ref. [14]. XPS simulations were performed with the Delta-Self-Consistent-Field ($\Delta$SCF) method. Here, for each carbon atom an excited core-hole calculation was performed, where the electron configuration of the respective pseudopotential was modified to [$1s^1$, $2s^2$, $2p^3$] to localise the core-hole. The additional valence electron was removed, and the net positive charge was compensated by a homogeneous background charge of −1.0 e to achieve net neutrality in the unit cell. XP spectra were broadened with a pseudo-Voigt function to capture instrumental and lifetime broadening effects.[15] For the SXPS modelling, a Gaussian full width half maximum (FWHM) of 0.40 eV and a Lorentzian FWHM of 0.46 eV were used. A global shift on the simulated spectrum is applied to align with experimental binding energies. For the Stone-Wales defect structure reported in Figure 3 of the main manuscript, XPS binding energies were calculated for all carbon atoms within the Stone-Wales defect and in the 1st, 2nd, 3rd, and 4th neighbour shell. XPS binding energies were calculated for 16 randomly placed carbon atoms for the metal-adsorbed ideal graphene sheet. The atoms considered in the XPS and NEXAFS simulations are indicated in colour in the Figure S10.

## 8. Details on the computed XPS and NEXAFS data: NEXAFS calculations

Simulations were carried out using the ELNES module[21] in CASTEP. A half core-hole was introduced into the pseudopotential for carbon by changing the electron configuration of the pseudopotential to [$1s^{1.5}$, $2s^2$, $2p^{2.5}$]. Each core-hole calculation requires a self-consistent total energy calculation followed by a band structure and dipole matrix element calculation to converge the unoccupied states. 1400, 900, and 2000 unoccupied states were considered, for the SW defect, the metal-adsorbed graphene and the free-standing ideal graphene, which was the limit achievable with the computational memory and time constraints. The total NEXAFS spectrum is produced by summing all single carbon species contributions and shifting them with the previously calculated XPS core binding energy. An energy-dependent broadening scheme was applied to account for the reduced lifetime at increasing excitation energies. The NEXAFS spectrum is divided into three ranges. The first range starts with the leading edge, spans the first 5 eV of the spectrum and is assigned a pseudo-Voigt FWHM of 0.75 eV and a 80%/20% Gaussian/Lorentzian ratio, while the third range starts from
15 eV above the leading edge and is assigned a FWHM of 2.0 eV and a 20%/80% G/L ratio. Both ranges are connected by an intermediate range, in which the FWHM and the G/L ratio change linearly. A global energy shift is applied to the final broadened NEXAFS spectrum to align with the experimental data. Due to the large computational cost of the NEXAFS spectra, only a subset of the atoms for which XP binding energies were calculated were considered in the NEXAFS simulations. Note the simulated spectra only consider contributions from atoms within the defect and should only be considered indicative of the source of features in the experimental spectra.

**Figure S10**

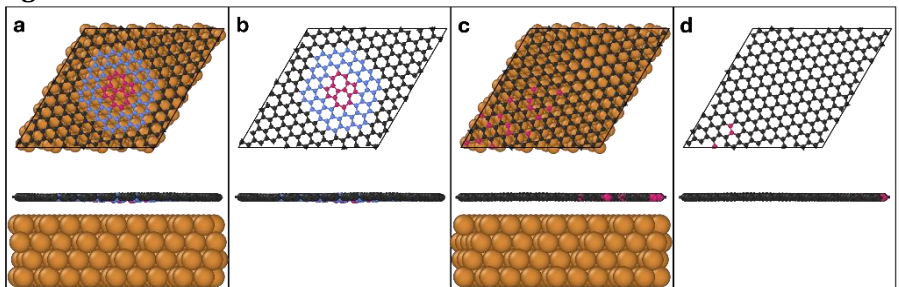

**Figure S10:** Top and side views of the calculated moiré pattern of graphene with a Stone-Wales defect (**a**) adsorbed on a Cu(111) surface and (**b**) freestanding. Top and side views of the calculated moiré pattern of ideal graphene (**c**) adsorbed on a Cu(111) surface and (**d**) freestanding. Carbon atoms coloured red were used in the XPS and NEXAFS calculations, coloured blue in the XPS calculations only and coloured black were not used in either spectroscopy calculation. Note that the NEXAFS calculations were performed with the bottom 2 Cu layers removed (not depicted) to maintain tractability with the increased computational cost of performing NEXAFS simulations. Visualisations were created with Ovito.[22]

## 9. Shifting of the binding energy and photon energy scales in the calculated spectra

In Figure 3, in the main article, the simulated spectra in panel **b** have been shifted so that the maximum intensity in the spectrum is the same as that measured experimentally for the SXPS data of the defective film. In panels **e,h** and **i** have been shifted in photon energy so that the graphene $\pi^*$ resonance matches that of the experimental data and have been rescaled in intensity such that the minima between the $\pi^*$ and $\sigma^*$ region overlaps for the defective and graphene theoretical data. The simulated XPS and NEXAFS spectra without an applied shift in binding / photon energy are shown in Figure S11. NB. in Figure 3**g** the spectra have been offset such that all spectra have identical intensity at a photon energy of 289 eV.

# Figure S11

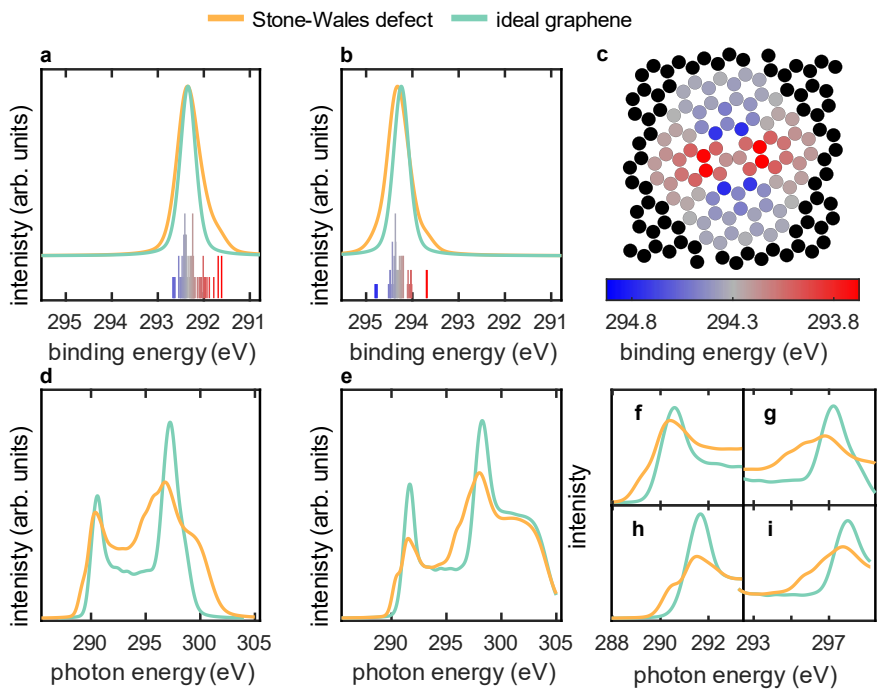

**Figure S11:** Computed XPS of a Stone-Wales defect embedded in a graphene mesh and the associated ideal graphene mesh (**a**) on Cu(111) and (**b**) free-standing, with a histogram of the theoretical binding energies of the Stone-Wales defect spectrum, without broadening, using the same colourscale as shown in panel (**c**). Also shown (**c**) is a top view depiction of the structure employed in the free-standing DFT calculations for the Stone-Wales defect, showing the binding energy of each calculated atom as per the indicated colourscale. Computed NEXAFS of a Stone-Wales defect embedded in a graphene mesh and the associated ideal graphene mesh (**d,f,g**) on Cu(111) and (**e,h,i**) free-standing. Note that these spectra are not offset to match the experimental data, but are shown with their calculated photon and binding energies.

## 10. Normal incidence X-ray standing waves

The normal incidence X-ray standing wave (NIXSW) technique[23] utilises the X-ray standing wave formed by the constructive / deconstructive interference between the incident and reflected waves close to the Bragg condition for a given Bragg reflection H=(h,k,l). The period of this standing wave matches that of the interplanar spacing $d_H$ between the Bragg diffraction planes[24]. The phase of the standing wave varies when the photon energy is scanned through the Bragg condition. This phase change results in the location of the maximum intensity of the standing wave with respect to the Bragg diffraction planes itself varying. When the phase is zero, the maximum intensity of the standing wave lies halfway between Bragg diffraction planes; when the phase is π the maximum intensity is coincident with the Bragg diffraction planes. In the case of face centred cubic crystal made up of a single element, the Bragg diffraction planes will be coincident with the real space atomic planes of the material. Any atom immersed in this standing wavefield will experience a varying electromagnetic field intensity as a function of its position between these diffraction planes. This varying field intensity will result in characteristic absorption profiles that can be monitored by the photoelectron intensity profile. The measured profile is then fitted uniquely, using dynamical diffraction theory,[25] by two dimensionless parameters[23]: the coherent fraction, $f_H$, and the coherent position, $p_H$. These, respectively, broadly correspond to the degree of order and the mean position of the absorber atoms relative to the Bragg diffraction planes. When the chosen Bragg diffraction plane is parallel with the surface plane the coherent position is related to the mean adsorption height of a species by:

$$h_H = (n + p_H) \cdot d_H, \tag{1}$$

where $n$ is an integer and relates to so called "modulo-$d$" ambiguity[26]. This ambiguity means that adsorption heights that differ by the interplanar spacing cannot be directly differentiated. However, in practice the correct value of $n$ can often be easily assigned as $d_H$ typically is in the order of ~2 Å, thus it is generally trivial to exclude adsorption heights that are unphysically low or high.

To model the experimentally measured coherent fraction and coherent position one must sum over all positions that the measured atoms take within the layer spacing defined by the Bragg diffraction plane. As such:

$$f_H \cdot \exp(i \cdot 2\pi p_H) = \sum_{n=1}^{n=m} f_n \cdot \exp(i \cdot 2\pi p_n), \tag{2}$$

where $f_n$ is the fraction of atoms that are at the coherent position, $p_n$ and $n = 1 \rightarrow m$ are all the possible positions that the species can take. If one considers a simple case, with an equal mixture of two adsorption sites, equation (2) becomes:

$$f_H \cdot \exp(i \cdot 2\pi p_H) = \frac{\exp(i \cdot 2\pi p_1)}{2} + \frac{\exp(i \cdot 2\pi p_2)}{2}. \tag{3}$$

For the special case where $p_1$ differs from $p_2$ by 0.5 then,

$$f_H \cdot \exp(i \cdot 2\pi p_H) = \frac{\exp(i \cdot 2\pi p_1)}{2} + \frac{\exp(i \cdot 2\pi (p_1+0.5))}{2}, \tag{4}$$

$$f_H \cdot \exp(i \cdot 2\pi p_H) = \frac{\exp(i \cdot 2\pi p_1)}{2} - \frac{\exp(i \cdot 2\pi p_1)}{2}, \tag{5}$$

$$f_H \cdot \exp(i \cdot 2\pi p_H) = 0, \tag{6}$$

$$f_H = 0. \tag{7}$$

Thus, if a species constitutes two primary adsorption heights that happen to differ by approximately half of the $d$-spacing related to the measured Bragg plane, then the expected measured coherent fraction would be very low.

## 11. Calibration of sample growth temperature

The various films presented in this work were grown in different sample preparation chambers, specifically an offline growth chamber at Diamond Light Source, the growth chamber directly mounted on the I09 beam line at Diamond Light Source, the growth chamber directly mounted on the FlexPES beam line at MAX IV and the growth chamber directly mounted on the nc-AFM system at the Technical University of Munich. On the two growth chambers at Diamond Light Source there was no thermocouple mounted close to the sample and, through the many sample preparations presented in this work, the methodology for mounting the samples onto the sample plates differed significantly. In short, there is significant uncertainty in the temperatures used for the growth of the samples at Diamond Light Source. However, the samples grown in the growth chamber directly mounted on the FlexPES beam line at MAX IV had a thermocouple directly mounted at the sample. Excluding all nc-AFM samples (Figure 3) and STM measurements of the molecular and graphene films shown in Figure 1 of the main article, every single sample presented in this work was measured by SXPS, thus by comparison of the SXPS data for the samples measured on the FlexPES beam line at MAX IV to the SXPS data measured for all other samples, we can obtain an estimate of the sample growth temperature. Specifically, as shown in Figure S12, the MAX IV SXPS data shows a clear linear variation in the binding energy of the maximum in intensity as a function of growth temperature. By fitting the linear variation observed in the MAX IV data we could estimate the growth temperature of the samples grown at Diamond Light Source by using the binding energy of the maximum intensity measured in their respective SXP spectra.

**Figure S12**

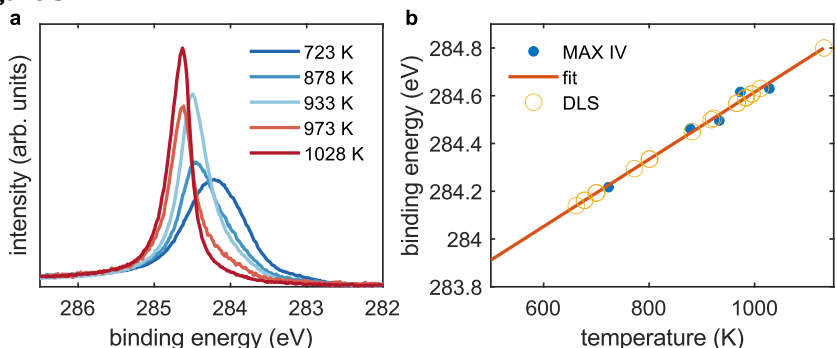

**Figure S12**: (**a**) XPS obtained of the samples grown in the sample growth chamber mounted directly onto the FlexPES beam line at MAX IV and (**b**) the binding energy of the maximum intensity plotted against their growth temperature that was measured directly by a thermocouple mounted onto the sample. The linear trend of the binding energy of the maximum intensity as a function of growth temperature is fitted with a straight line and used to estimate the growth temperature of the samples grown at Diamond Light Source (DLS).

**Figures S13-S19**

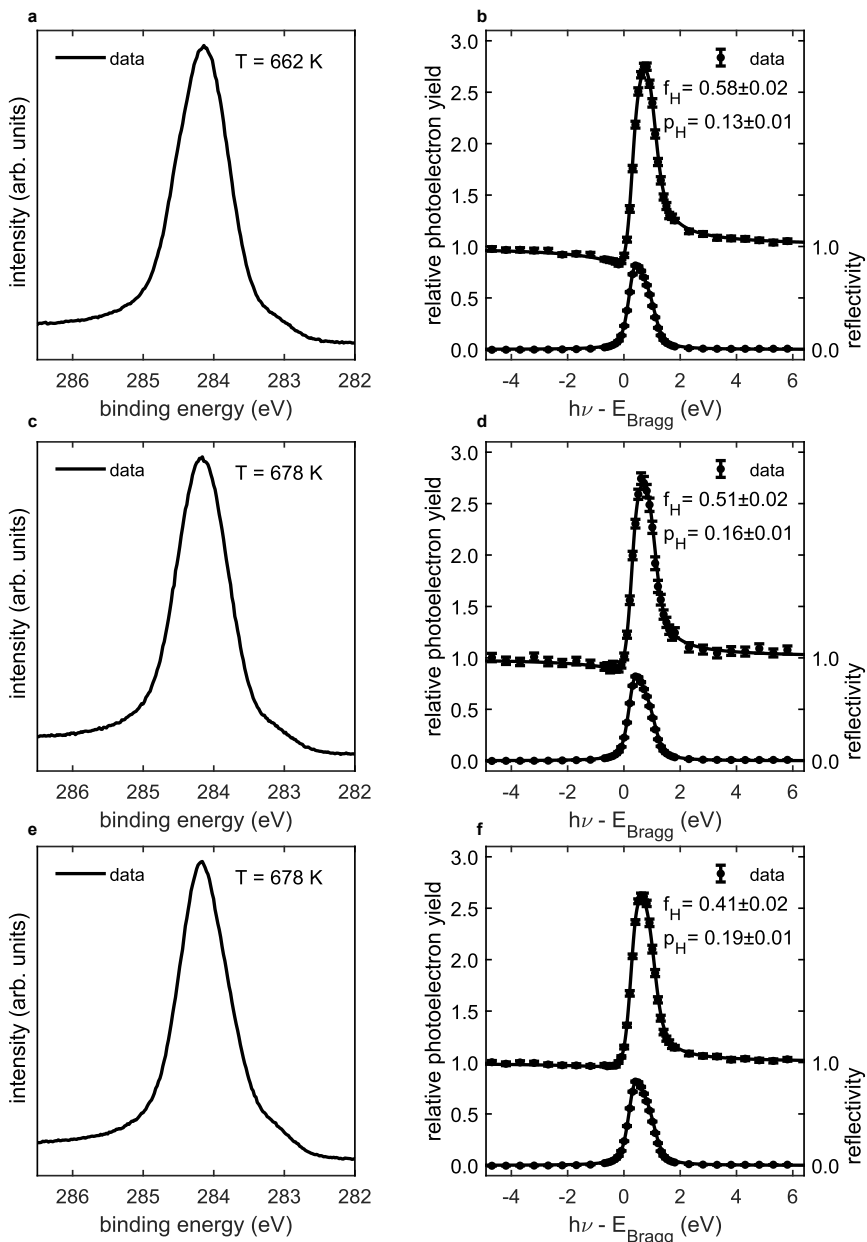

**Figure S13:** (**a,c,e**) Experimental C 1s SXPS and (**b,d,f**) NIXSW data for films grown on Cu(111) held at the indicated estimated temperature under exposure to azupyrene. The indicated coherent fractions and coherent positions, obtained from fitting the NIXSW form the data shown in Figure 4b,c in the main manuscript.

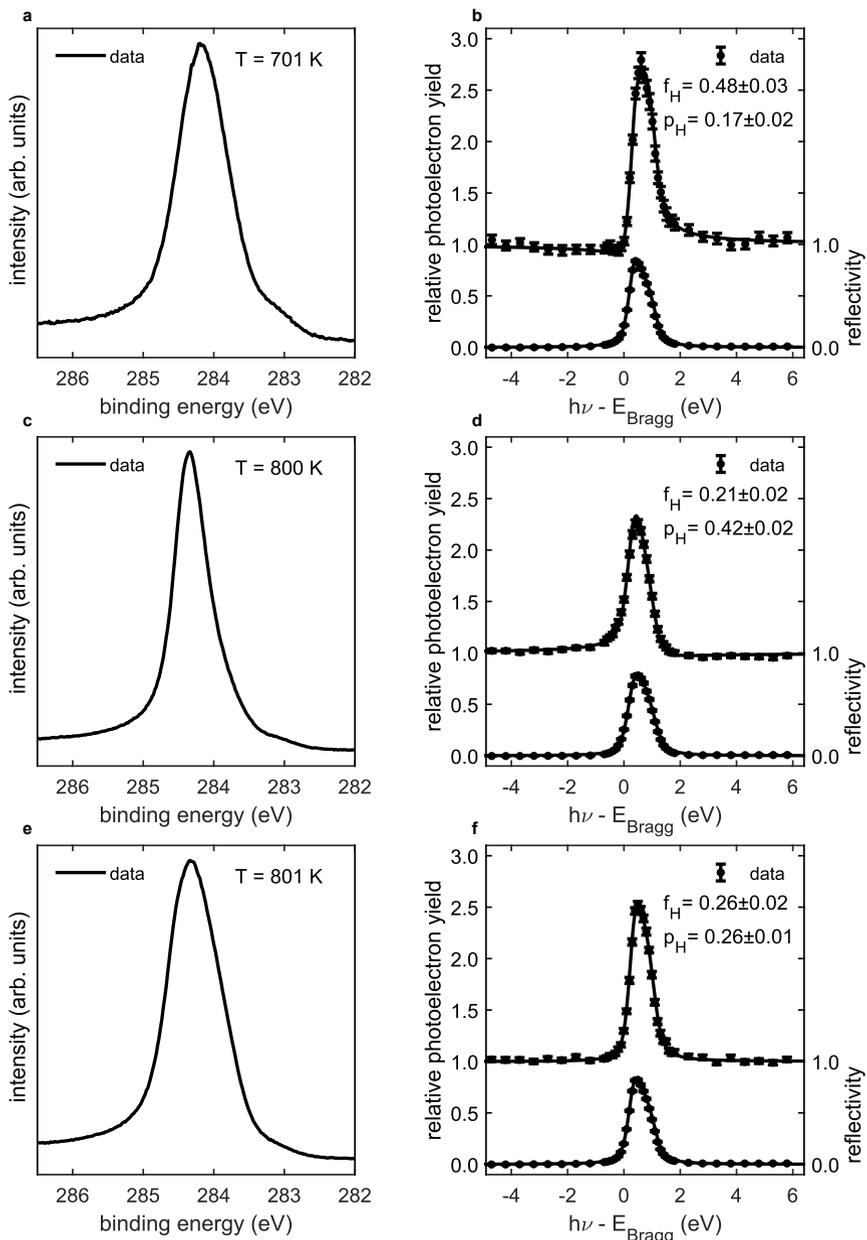

**Figure S14:** (**a**,**c**,**e**) Experimental C 1s SXPS and (**b**,**d**,**f**) NIXSW data for films grown on Cu(111) held at the indicated estimated temperature under exposure to azupyrene. The indicated coherent fractions and coherent positions, obtained from fitting the NIXSW form the data shown in Figure 4b,c in the main manuscript.

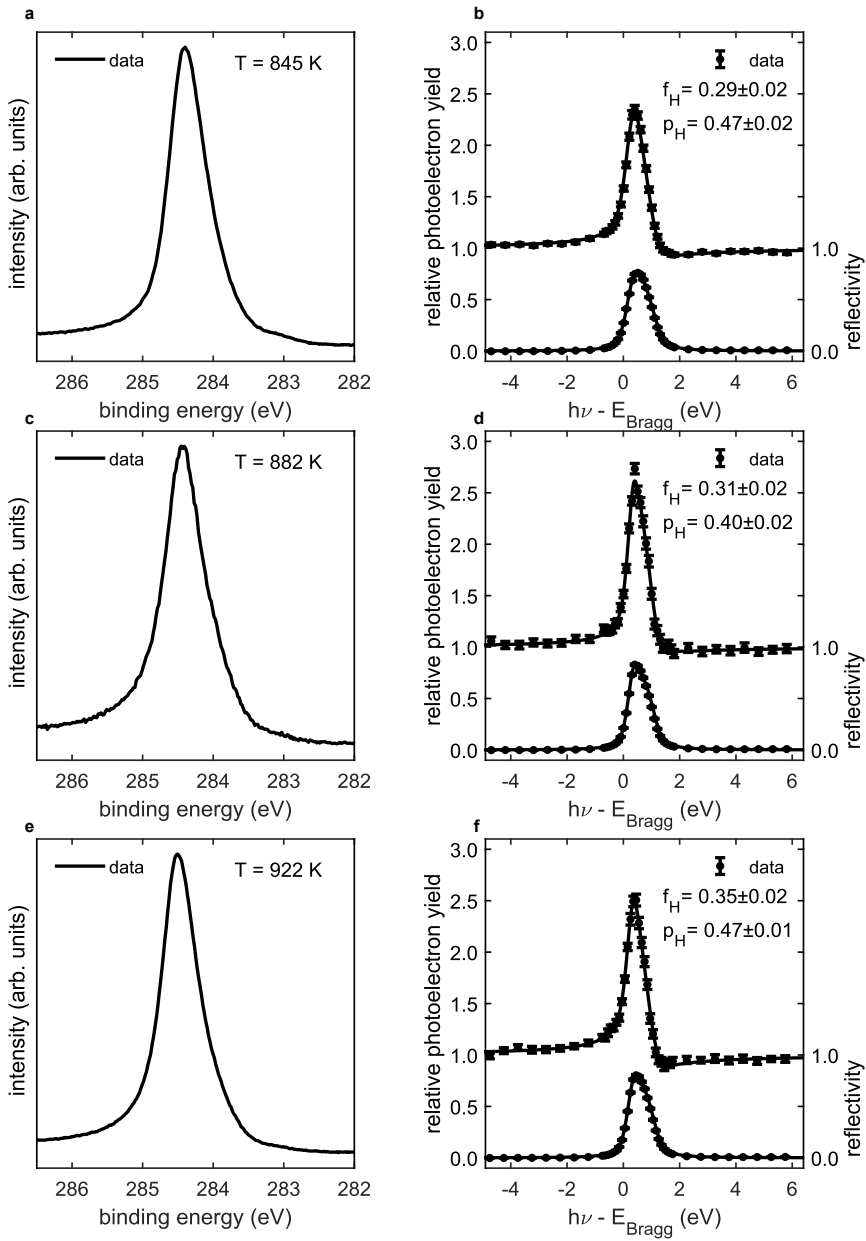

**Figure S15:** (**a**,**c**,**e**) Experimental C 1s SXPS and (**b**,**d**,**f**) NIXSW data for films grown on Cu(111) held at the indicated estimated temperature under exposure to azupyrene. The indicated coherent fractions and coherent positions, obtained from fitting the NIXSW form the data shown in Figure 4b,c in the main manuscript.

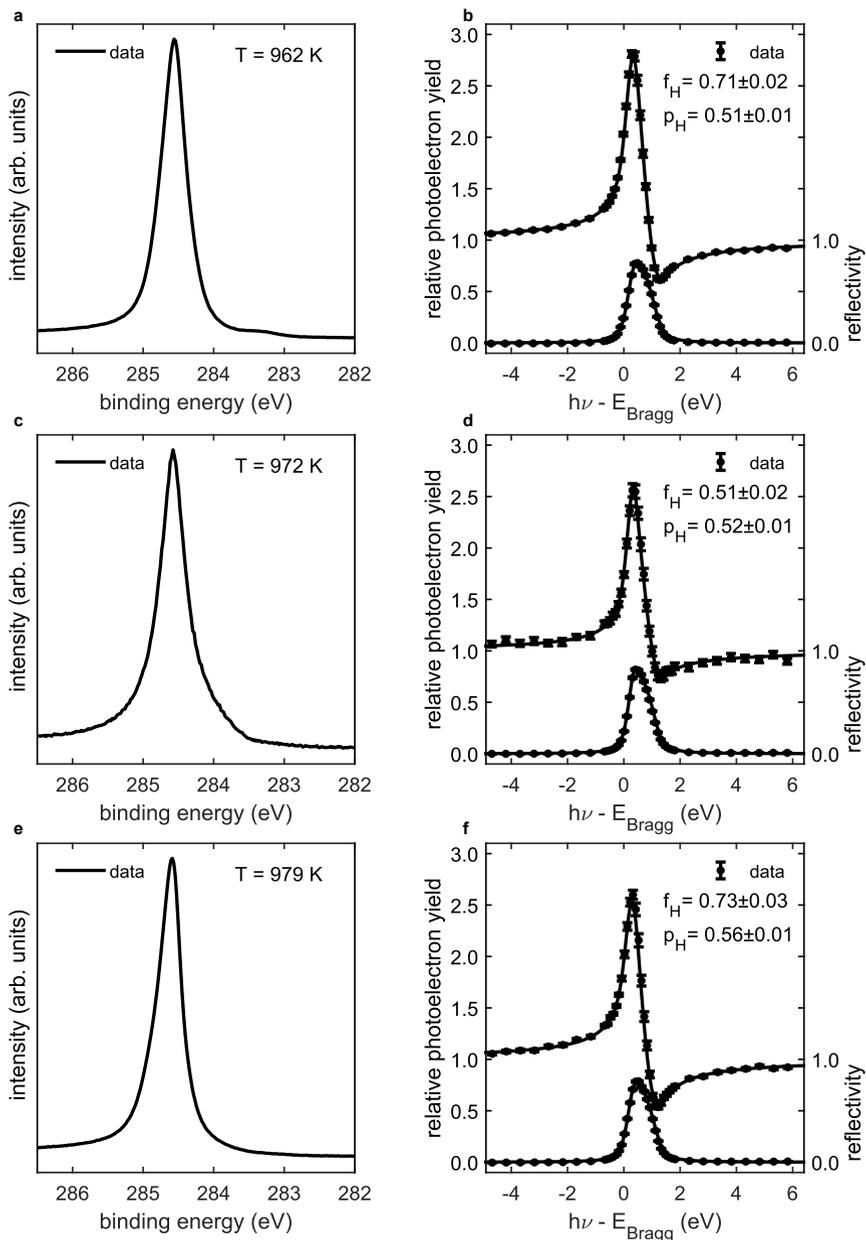

**Figure S16:** (**a**,**c**,**e**) Experimental C 1s SXPS and (**b**,**d**,**f**) NIXSW data for films grown on Cu(111) held at the indicated estimated temperature under exposure to azupyrene. The indicated coherent fractions and coherent positions, obtained from fitting the NIXSW form the data shown in Figure 4b,c in the main manuscript.

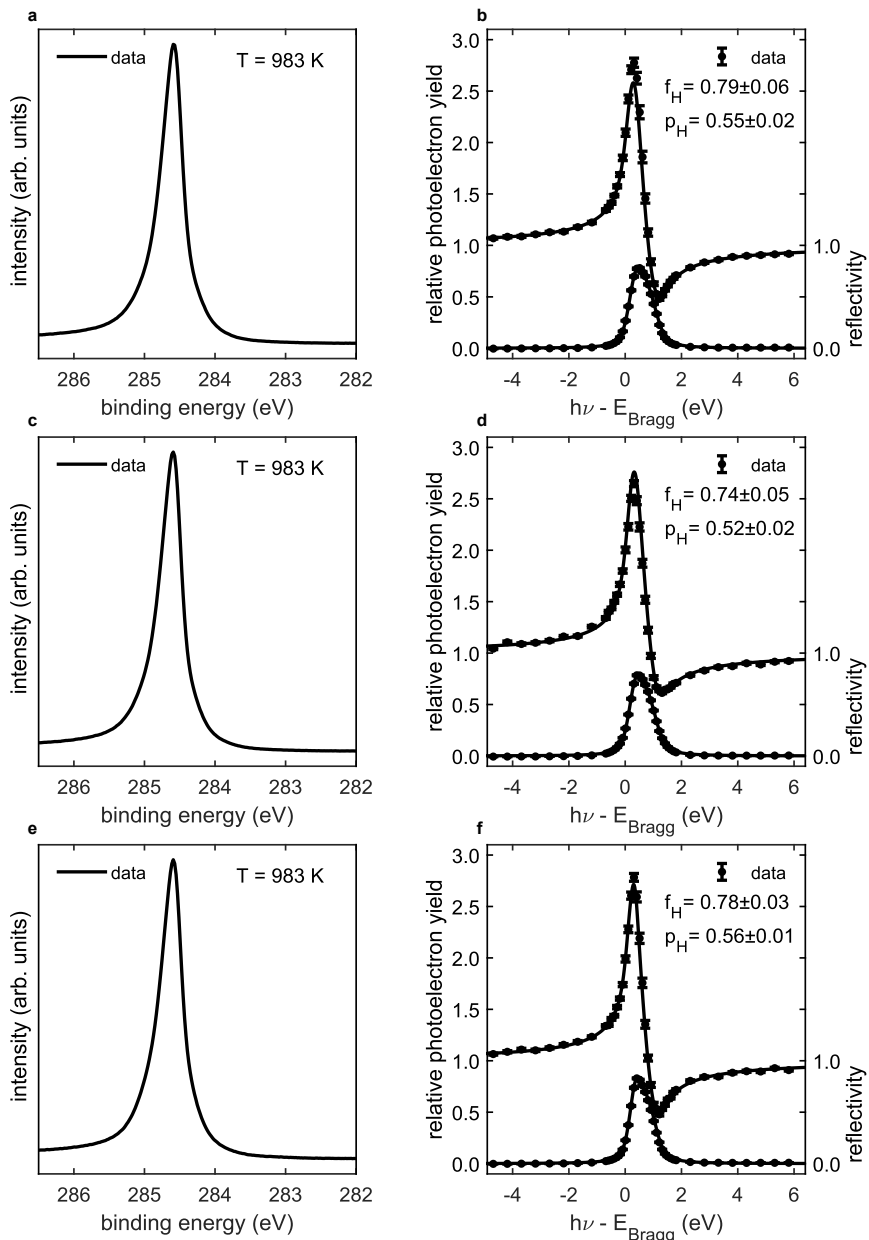

**Figure S17:** (**a**,**c**,**e**) Experimental C 1s SXPS and (**b**,**d**,**f**) NIXSW data for films grown on Cu(111) held at the indicated estimated temperature under exposure to azupyrene. The indicated coherent fractions and coherent positions, obtained from fitting the NIXSW form the data shown in Figure 4b,c in the main manuscript.

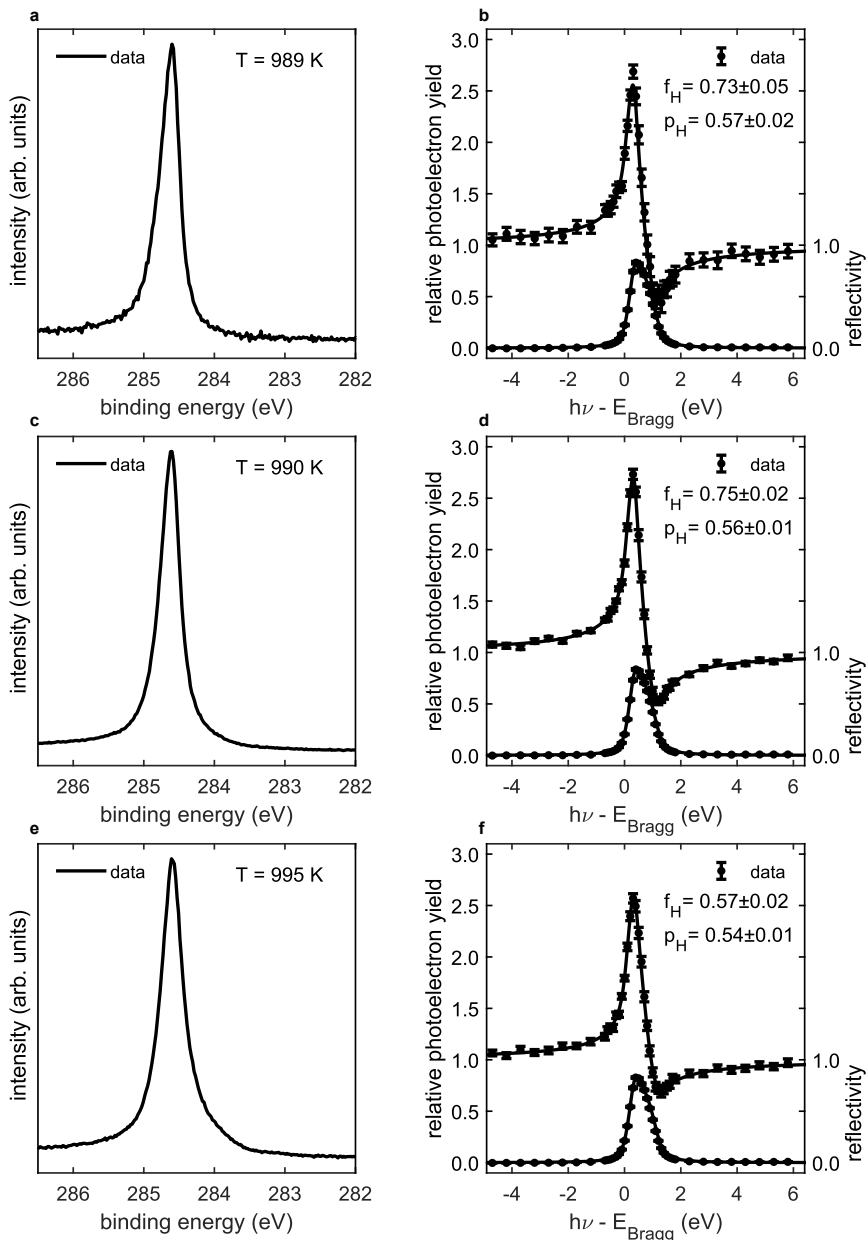

**Figure S18:** (**a**) Experimental C 1s SXPS and (**b**) NIXSW data for films grown on Cu(111) held at the indicated estimated temperature under exposure to azupyrene. The indicated coherent fractions and coherent positions, obtained from fitting the NIXSW form the data shown in Figure 4b,c in the main manuscript.

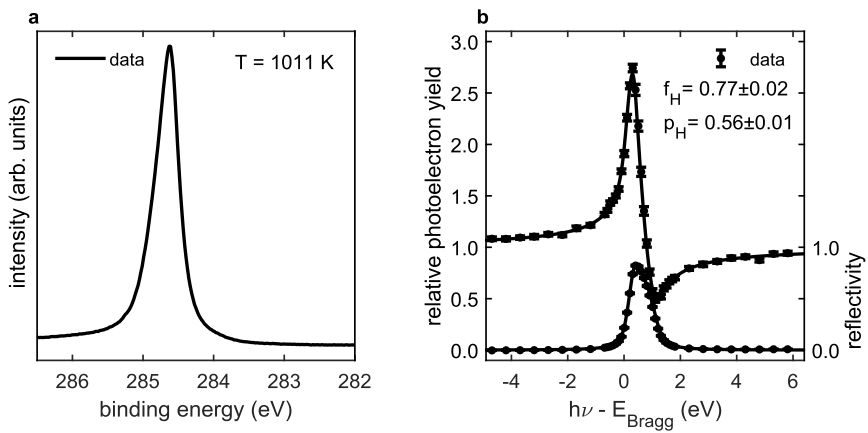

**Figure S19:** (a,c,e) Experimental C 1s SXPS and (b,d,f) NIXSW data for films grown on Cu(111) held at the indicated estimated temperature under exposure to azupyrene. The indicated coherent fractions and coherent positions, obtained from fitting the NIXSW form the data shown in Figure 4b,c in the main manuscript.

## 12. Further details on the ADF-STEM measurements

The enumerated frequency of the identified 4- to 8-membered rings can be found in Table S1. Small bright protrusions were observed on occasion, e.g. the circled areas in Figure S22, which may indicate small clusters of large Z atoms that have remained on defect sites. These protrusions are very small, potentially even single atoms. It is well established that the high energy electron beam used in TEM measurement can induce 5- and 7-membered defects, such as those observed in this study, though it is more likely to heal such a defect than create one.[27] As such, we performed the measurements in a low energy mode with a reduced electron flux to reduce the probability of defect introduction. In this regime, the probabilistic rate of beam induced 5-/7-membered ring formation would yield one ring per scan. Considering the rate of C ejection observed by Meyer et al.[28], the expected defect rate induced by the beam in our AC-STEM measurements would be ~1 defect per image.

**Figures S20-24**

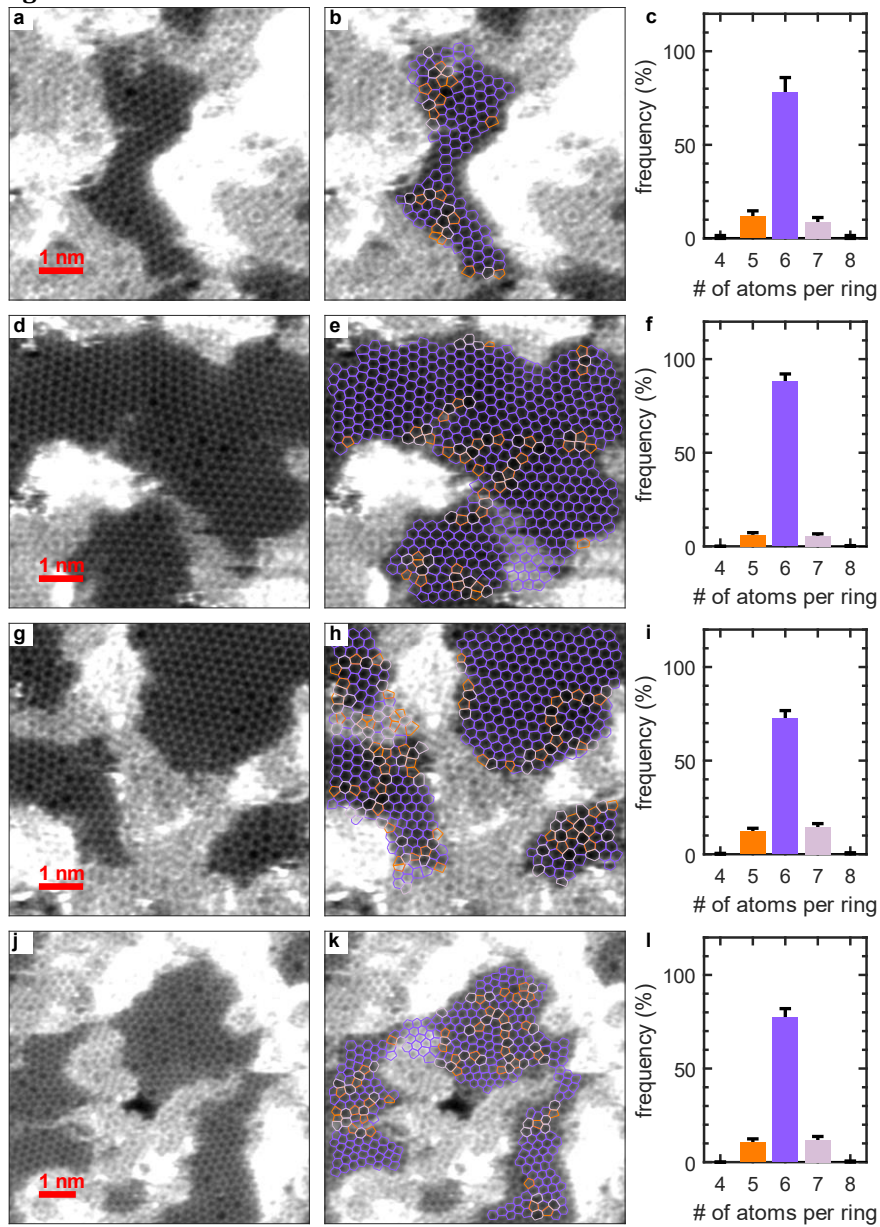

**Figure S20:** (**a-b**,**d-e**,**g-h**,**j-k**) Atomically resolved AC-STEM measurements of a defective film. Overlayed in panels (**b**,**e**,**h**,**k**) are the C-C bonds, 5-membered rings are dark orange, 7-membered are light purple and 6-membered are dark purple. Also shown in panels (**c**,**f**,**i**,**l**) are the histograms of the relative frequency of 4- to 8-membered rings.

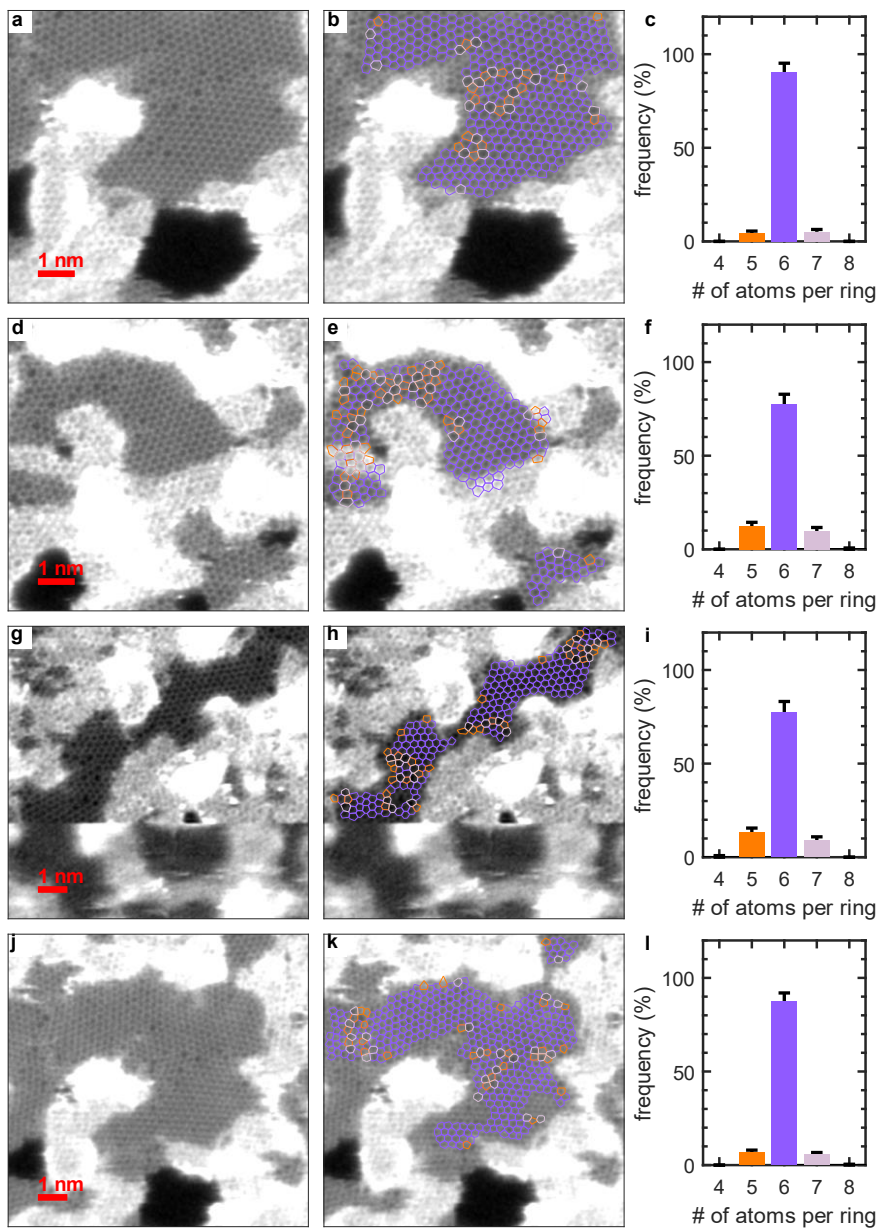

**Figure S21:** (**a-b**,**d-e**,**g-h**,**j-k**) Atomically resolved AC-STEM measurements of a defective film. Overlayed in panels (**b**,**e**,**h**,**k**) are the C-C bonds, 5-membered rings are dark orange, 7-membered are light purple and 6-membered are dark purple. Also shown in panels (**c**,**f**,**i**,**l**) are the histograms of the relative frequency of 4- to 8-membered rings.

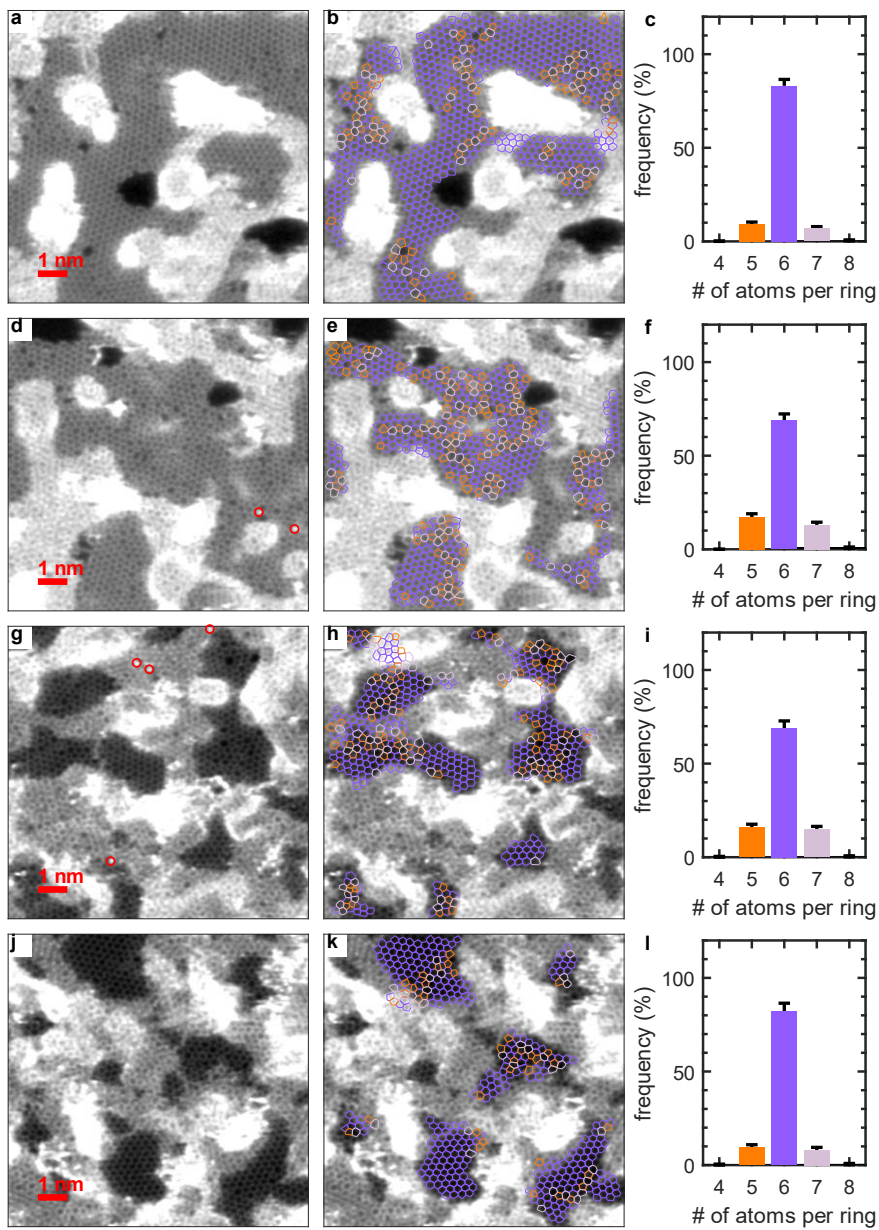

**Figure S22:** (**a-b**,**d-e**,**g-h**,**j-k**) Atomically resolved AC-STEM measurements of a defective film. Red circles highlight protrusions, that may relate to single metal atoms. Overlayed in panels (**b**,**e**,**h**,**k**) are the C-C bonds, 5-membered rings are dark orange, 7-membered are light purple and 6-membered are dark purple. Also shown in panels (**c**,**f**,**i**,**l**) are the histograms of the relative frequency of 4- to 8-membered rings.

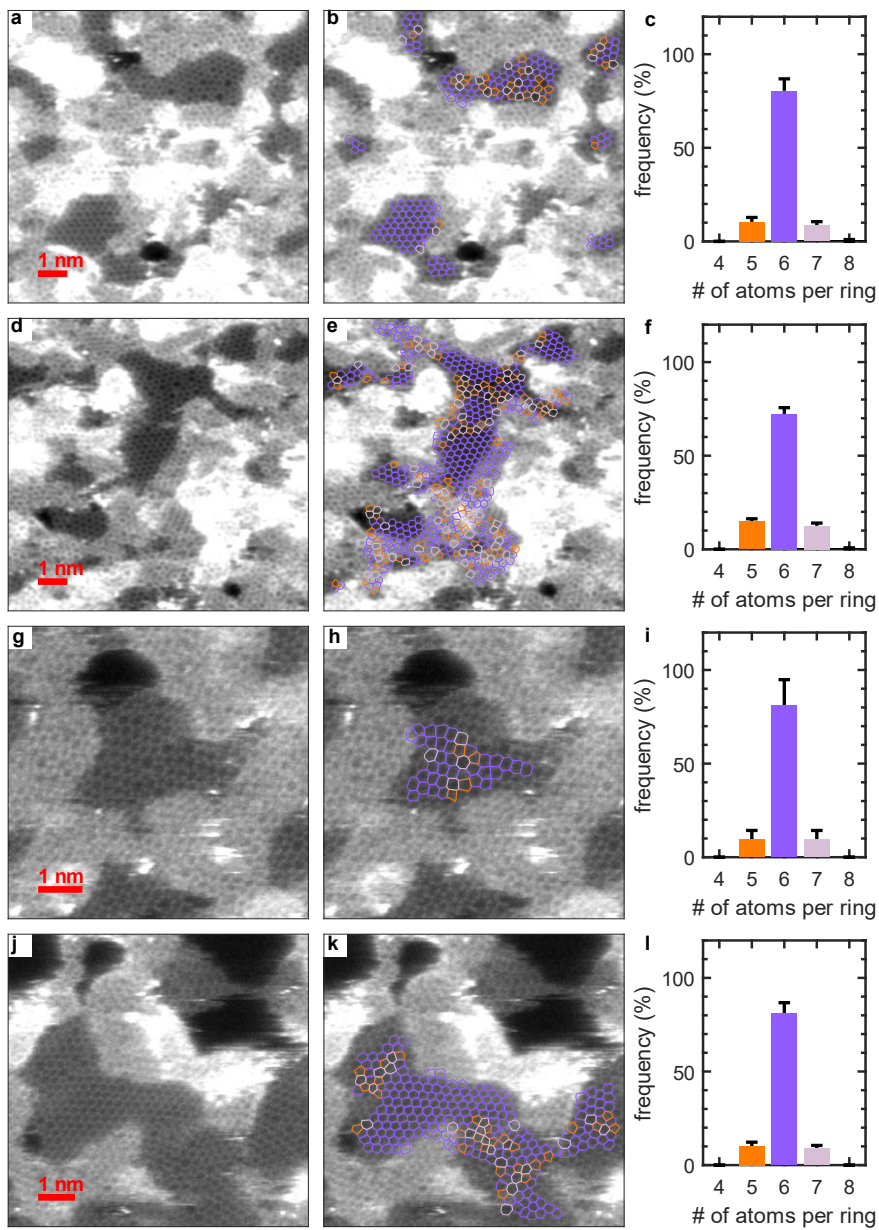

**Figure S23:** (**a-b**,**d-e**,**g-h**,**j-k**) Atomically resolved AC-STEM measurements of a defective film. Overlayed in panels (**b**,**e**,**h**,**k**) are the C-C bonds, 5-membered rings are dark orange, 7-membered are light purple and 6-membered are dark purple. Also shown in panels (**c**,**f**,**i**,**l**) are the histograms of the relative frequency of 4- to 8-membered rings.

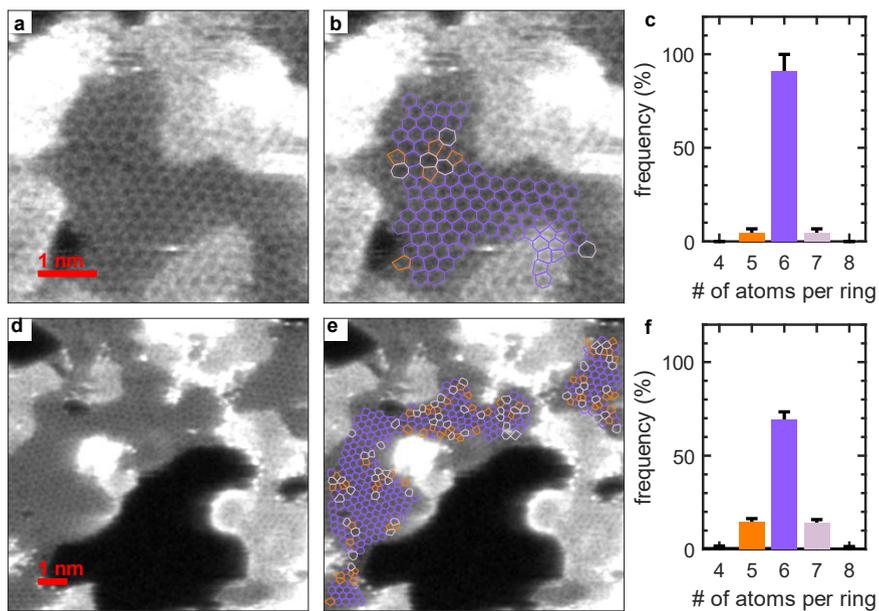

**Figure S24:** (**a-b**,**d-e**) Atomically resolved AC-STEM measurements of a defective film. Overlayed in panels (**b**,**e**) are the C-C bonds, 5-membered rings are dark orange, 7-membered are light purple and 6-membered are dark purple. Also shown in panels (**c**,**f**) are the histograms of the relative frequency of 4- to 8-membered rings.